\documentclass[journal]{IEEEtran}

\usepackage{cite,graphicx,subfigure,hyperref,amsmath,amsthm,amsfonts,array,xcolor} 
\usepackage{diagbox}

\newlength{\Lpr}
\newsavebox{\Bpr}

\newcommand{\D}[1]{\ensuremath{\displaystyle #1}}

\newcommand{\V}[1]{\mbox{\boldmath$#1$\unboldmath}}

\newcommand{\bdm}{\begin{displaymath}}
\newcommand{\edm}{\end{displaymath}}

\newcommand{\be}[1]{\begin{equation} \label{#1}}
\newcommand{\ee}{\end{equation}}

\newcommand{\bae}[3]{
\begin{equation} \label{#1}
\renewcommand{\arraystretch}{#2}
\begin{array}{#3}}

\newcommand{\eae}{\end{array}\end{equation}}

\newcommand{\baen}[2]{
\begin{displaymath} 
\renewcommand{\arraystretch}{#1}
\begin{array}{#2}}

\newcommand{\eaen}{\end{array}\end{displaymath}}

\newcommand{\DefLetter}[4]{
\newcommand{#1}{\ensuremath{\V{#2}}} 
\newcommand{#3}{\ensuremath{\V{#4}}} 
}

\DefLetter{\vzer}{0}{\mzer}{0}
\DefLetter{\vone}{1}{\mone}{1}
\DefLetter{\va}{a}{\ma}{A}
\DefLetter{\vb}{b}{\mb}{B}
\DefLetter{\vc}{c}{\mc}{C}
\DefLetter{\vd}{d}{\md}{D}
\DefLetter{\ve}{e}{\me}{E}
\DefLetter{\vf}{f}{\mf}{F}
\DefLetter{\vg}{g}{\mg}{G}
\DefLetter{\vh}{h}{\mh}{H}
\DefLetter{\vi}{i}{\mi}{I}
\DefLetter{\vj}{j}{\mj}{J}
\DefLetter{\vk}{k}{\mk}{K}
\DefLetter{\vl}{l}{\ml}{L}
\DefLetter{\vm}{m}{\mm}{M}
\DefLetter{\vn}{n}{\mn}{N}
\DefLetter{\vpr}{p}{\mpr}{P}
\DefLetter{\vq}{q}{\mq}{Q}
\DefLetter{\vr}{r}{\mr}{R}
\DefLetter{\vs}{s}{\ms}{S}
\DefLetter{\vt}{t}{\mt}{T}
\DefLetter{\vur}{u}{\mur}{U}
\DefLetter{\vv}{v}{\mv}{V}
\DefLetter{\vw}{w}{\mw}{W}
\DefLetter{\vx}{x}{\mx}{X}
\DefLetter{\vy}{y}{\my}{Y}
\DefLetter{\vz}{z}{\mz}{Z}

\DefLetter{\vdel}{\delta}{\mdel}{\Delta}
\DefLetter{\vphi}{\phi}{\mphi}{\Phi}
\DefLetter{\vpsi}{\psi}{\mpsi}{\Psi}
\DefLetter{\vrho}{\rho}{\mrho}{\Lambda}
\DefLetter{\vxi}{\xi}{\mxi}{\Xi}

\DefLetter{\valpha}{\alpha}{\malpha}{\Alpha}
\DefLetter{\vbeta}{\beta}{\mbeta}{\Beta}
\DefLetter{\vlam}{\lambda}{\mlam}{\Lambda}
\DefLetter{\vsig}{\sigma}{\msig}{\Sigma}
\DefLetter{\vtau}{\tau}{\mtau}{\tau}
\DefLetter{\vtheta}{\theta}{\mtheta}{\Theta}
\DefLetter{\vome}{\omega}{\mome}{\Omega}
\DefLetter{\vzero}{0}{\mzero}{0}
\DefLetter{\vgam}{\gamma}{\mgam}{\Gamma}
\DefLetter{\veps}{\epsilon}{\meps}{\Epsilon}
\DefLetter{\veta}{\eta}{\meta}{\Eta}

\newcommand{\DefFuncLetter}[2]{
\newcommand{#1}{\ensuremath{{#2}}} 
}

\DefFuncLetter{\Fzer}{0}
\DefFuncLetter{\Fa}{a}
\DefFuncLetter{\FA}{A}
\DefFuncLetter{\Fb}{b}
\DefFuncLetter{\Fc}{c}
\DefFuncLetter{\FC}{C}
\DefFuncLetter{\Fd}{d}
\DefFuncLetter{\Fe}{e}
\DefFuncLetter{\Ff}{f}
\DefFuncLetter{\Fg}{g}
\DefFuncLetter{\FG}{G}
\DefFuncLetter{\Fh}{h}
\DefFuncLetter{\FH}{H}
\DefFuncLetter{\Fi}{i}
\DefFuncLetter{\Fk}{k}
\DefFuncLetter{\Fl}{l}
\DefFuncLetter{\FL}{L}
\DefFuncLetter{\Fm}{m}
\DefFuncLetter{\Fn}{n}
\DefFuncLetter{\Fnr}{n}
\DefFuncLetter{\FN}{N}
\DefFuncLetter{\Fo}{o}
\DefFuncLetter{\FO}{O}
\DefFuncLetter{\Fpr}{p}
\DefFuncLetter{\FPr}{P}
\DefFuncLetter{\Fq}{q}
\DefFuncLetter{\Fr}{r}
\DefFuncLetter{\Fs}{s}
\DefFuncLetter{\FS}{S}
\DefFuncLetter{\Ft}{t}
\DefFuncLetter{\FT}{T}
\DefFuncLetter{\Fu}{u}
\DefFuncLetter{\FU}{U}
\DefFuncLetter{\Fv}{v}
\DefFuncLetter{\Fw}{w}
\DefFuncLetter{\FW}{W}
\DefFuncLetter{\Fx}{x}
\DefFuncLetter{\Fy}{y}
\DefFuncLetter{\FY}{Y}
\DefFuncLetter{\Fz}{z}
\DefFuncLetter{\FZ}{Z}
\DefFuncLetter{\Falp}{\alpha}
\DefFuncLetter{\Fbet}{\beta}
\DefFuncLetter{\Fchi}{\chi}
\DefFuncLetter{\Fdel}{\delta}
\DefFuncLetter{\Fzet}{\zeta}
\DefFuncLetter{\FEps}{\Epsilon}
\DefFuncLetter{\Feta}{\eta}
\DefFuncLetter{\Fphi}{\phi}
\DefFuncLetter{\FPhi}{\Phi}
\DefFuncLetter{\Fpsi}{\psi}
\DefFuncLetter{\FPsi}{\Psi}
\DefFuncLetter{\Fgam}{\gamma}
\DefFuncLetter{\FGam}{\Gamma}
\DefFuncLetter{\Flam}{\lambda}
\DefFuncLetter{\FLam}{\Lambda}
\DefFuncLetter{\Fmur}{\mu}
\DefFuncLetter{\Fsig}{\sigma}
\DefFuncLetter{\Ftau}{\tau}
\DefFuncLetter{\Fome}{\omega}
\DefFuncLetter{\Feps}{\epsilon}
\DefFuncLetter{\Fthe}{\theta}
\DefFuncLetter{\Fvar}{\vartheta}

\DefFuncLetter{\FB}{B}
\DefFuncLetter{\FD}{D}
\DefFuncLetter{\FE}{E}
\DefFuncLetter{\FF}{F}
\DefFuncLetter{\FI}{I}
\DefFuncLetter{\FJ}{J}
\DefFuncLetter{\FM}{M}
\DefFuncLetter{\FR}{R}
\DefFuncLetter{\FV}{V}
\DefFuncLetter{\FX}{X}

\newcommand{\DefCalLetter}[2]{
\newcommand{#1}{\ensuremath{\mathcal{#2}}} 
}
\DefCalLetter{\CC}{C}
\DefCalLetter{\CD}{D}

\DefCalLetter{\CS}{S}
\DefCalLetter{\CV}{V}

\newcommand{\DefSubLetter}[2]{
\newcommand{#1}{\mathrm{#2}} 
}

\DefSubLetter{\slzer}{0}
\DefSubLetter{\sla}{a}
\DefSubLetter{\slA}{A}
\DefSubLetter{\slb}{b}
\DefSubLetter{\slB}{B}
\DefSubLetter{\slc}{c}
\DefSubLetter{\slC}{C}
\DefSubLetter{\sld}{d}
\DefSubLetter{\slD}{D}
\DefSubLetter{\sle}{e}
\DefSubLetter{\slE}{E}
\DefSubLetter{\slf}{f}
\DefSubLetter{\slF}{F}
\DefSubLetter{\slg}{g}
\DefSubLetter{\slG}{G}
\DefSubLetter{\slh}{h}
\DefSubLetter{\slH}{H}
\DefSubLetter{\sli}{i}
\DefSubLetter{\slI}{I}
\DefSubLetter{\slk}{k}
\DefSubLetter{\sll}{l}
\DefSubLetter{\slL}{L}
\DefSubLetter{\slm}{m}
\DefSubLetter{\slM}{M}
\DefSubLetter{\sln}{n}
\DefSubLetter{\slnr}{n}
\DefSubLetter{\slN}{N}
\DefSubLetter{\slo}{o}
\DefSubLetter{\slp}{p}
\DefSubLetter{\slP}{P}
\DefSubLetter{\slq}{q}
\DefSubLetter{\slQ}{Q}
\DefSubLetter{\slr}{r}
\DefSubLetter{\slR}{R}
\DefSubLetter{\sls}{s}
\DefSubLetter{\slS}{S}
\DefSubLetter{\slt}{t}
\DefSubLetter{\slT}{T}
\DefSubLetter{\slu}{u}
\DefSubLetter{\slU}{U}
\DefSubLetter{\slv}{v}
\DefSubLetter{\slw}{w}
\DefSubLetter{\slW}{W}
\DefSubLetter{\slx}{x}
\DefSubLetter{\slX}{X}
\DefSubLetter{\sly}{y}
\DefSubLetter{\slY}{Y}
\DefSubLetter{\slz}{z}
\DefSubLetter{\slZ}{Z}

\DefSubLetter{\slalp}{\alpha}
\DefSubLetter{\slbet}{\beta}
\DefSubLetter{\sldel}{\delta}
\DefSubLetter{\slDel}{\Delta}
\DefSubLetter{\sleps}{\epsilon}
\DefSubLetter{\slgam}{\gamma}
\DefSubLetter{\slphi}{\phi}
\DefSubLetter{\sltau}{\tau}
\DefSubLetter{\slxi}{\xi}
\DefSubLetter{\slthe}{\theta}

\newcommand{\fab}{\ensuremath{[f_A,f_B]}}

\newcommand{\mdim}[2]{${#1\times #2}$}

\newcommand{\rangect}[2]{${#1=0,}$ $1,\ldots,$ $#2$}
\newcommand{\range}[2]{${#1=1,\,2,\ldots}$, $#2$}

\newcommand{\pxw}{\mpr(f)}

\newcommand{\per}{\widehat\mpr_0(f_r)}

\newcommand{\ex}[1]{\mathcal E \{#1\}}

\newcommand{\gridD}{
\color{blue}
\linethickness{0.2mm}
\multiput(0,0)(0,1){11}{\line(1,0){10}}
\multiput(0,0)(1,0){11}{\line(0,1){10}}

\linethickness{0.05mm}
\multiput(0,0)(0,0.5){21}{\line(1,0){10}}
\multiput(0,0)(0.5,0){21}{\line(0,1){10}}

\linethickness{0.02mm}
\multiput(0,0)(0,0.1){101}{\line(1,0){10}}
\multiput(0,0)(0.1,0){101}{\line(0,1){10}}

\put(-0.09,-0.3){0} \put(0.91,-0.3){1} \put(1.91,-0.3){2} \
\put(2.91,-0.3){3} \put(3.91,-0.3){4} \put(4.91,-0.3){5} \
\put(5.91,-0.3){6} \put(6.91,-0.3){7} \put(7.91,-0.3){8} \
\put(8.91,-0.3){9} \put(9.91,-0.3){10}

\put(-0.25,-0.1){0} \put(-0.25,0.9){1} \put(-0.25,1.9){2} \
\put(-0.25,2.9){3} \put(-0.25,3.9){4} \put(-0.25,4.9){5} \
\put(-0.25,5.9){6} \put(-0.25,6.9){7} \put(-0.25,7.9){8} \
\put(-0.25,8.9){9} \put(-0.25,9.9){10}
}

\newcommand{\Fig}[3]{
\begin{figure}
\setlength{\unitlength}{1cm}
\centering{
\fbox{
\begin{picture}(8.5,5.5)
\thicklines
#1
\ifnum#3=1
\gridD
\fi
\end{picture}
}
}
\caption{#2}
\end{figure}
}

\newcommand{\dispkeys}{0}
\newcommand{\sref}[1]{
\ifnum\dispkeys=1
#1
\fi
\ref{#1}}

\allowdisplaybreaks[4]

\begin{document}

\title{Wideband Subspace Estimation Through Projection Matrix Approximation}

\author{J. Selva 
 \thanks{Submitted to the IEEE Transactions on Signal Processing}

}

\maketitle

\markboth{}{}
\begin{abstract}

  In this paper, we present a wideband subspace estimation method that  characterizes the signal subspace through its orthogonal projection matrix at each frequency.  Fundamentally, the method models this projection matrix as a function of frequency that can be approximated by a polynomial. It provides two improvements: a reduction in the number of parameters required to represent the signal subspace along a given frequency band and a quality improvement in wideband direction-of-arrival (DOA) estimators such as Incoherent Multiple Signal Classification (IC-MUSIC) and Modified Test of Orthogonality of Projected Subspaces (MTOPS). In rough terms, the method fits a polynomial to a set of projection matrix estimates, obtained at a set of frequencies, and then uses the polynomial as a representation of the signal subspace. The paper includes the derivation of asymptotic bounds for the bias and root-mean-square (RMS) error of the projection matrix estimate and a numerical assessment of the method and its combination with the previous two DOA estimators.
  
\end{abstract}

\section{Introduction}

In array processing, the estimation of the subspace spanned by several sources is a fundamental step in DOA estimation  \cite[Ch. 9]{VanTreesP4}. This estimation relies on the so-called narrowband condition, i.e, the array response must be constant in the spectral band covered by the sources.
In practice, however, this condition is often unrealistic in applications involving high data rates or acoustic or seismic signals, in which the array response varies with frequency significantly. In the literature on DOA estimation, these cases are classified as ``wideband'' and addressed by dividing the signals' band into bins in which the array response is approximately constant.
The problem is then the way the data from all bins should be combined in order to produce a single set of DOA estimates. In the literature, there are fundamentally two approaches for this combination. The first is the coherent approach in which the data from all bins are linearly combined \cite{Wang85,Valaee95,Yasar08,Zeng10}, and the second is the incoherent approach in which the combination is performed by other means, such as averaging the DOA narrowband estimates from each bin or adding up the bin pseudo-spectrum functions (as is done in IC-MUSIC),  \cite{Wax84}. Additionally, there exist other ways to process the bin data derived from general principles such as Maximum Likelihood (ML) \cite{Doron93,Yip02,Chen02,Yip08,Selva18b}, polynomial matrix decompositions  \cite{McWirther07,Alrmah11,Weiss13,Redif17}, or group sparsity \cite{Boufounos11,Shen14,Shen15}.

There is a relevant feature in this estimation problem that seems to be overlooked in the literature, which is the fact that the array response varies smoothly along the band covered by the impinging signals, and the same happens with the corresponding signal subspace. This smooth variation is a consequence of two basic facts. First, for any sensor array, there is an upper bound for the distance among sensors that can be measured as a delay $\tau_{max}$ at the propagation speed. And second, if the array is viewed as a vector Linear Time-Invariant system, then its spectrum is a band-limited function of bandwidth at most $\tau_{max}$. Thus, the array response variation with frequency is smooth (a band-limited function) and we may even bound any of its derivatives using standard results; (see Bernstein's inequality \cite[Th. 6.7]{Higgins96}). Additionally, if we consider a number of impinging waves from different DOAs, then the subspace spanned by the array response spectrum varies smoothly with frequency. To be more precise, the orthogonal projection matrix of the subspace varies smoothly with frequency and can be approximated in any given frequency band using a polynomial. This is a consequence of the fact that this projection matrix is analytic \cite[Ch. 2, Th. 1.10]{Kato95} and Weiertrass theorem \cite[Th. 2.4.1]{Phillips03}.

The purpose of this paper is to present a method for exploiting this last subspace smoothness in order to improve wideband subspace estimation. The method is based on a polynomial approximation and its input is the set of projection matrices that is usually computed in the binning approach.  Fundamentally, the method consists of fitting a polynomial to these last matrices using weighted least squares, and then evaluating the polynomial at any frequency for obtaining an approximate projection matrix. 

The next section is a extended introduction in which we justify the smooth variation of the signal subspace, outline the proposed method, and describe the organization of the paper. 

\subsection{Notation and main symbols}

We use the following notation and basic concepts:

\begin{itemize}
\item We write vectors in lower case ($\va$, $\vx$) and matrices in upper case, ($\ma$, $\mx$).
\item $\mi_M$ is an identity matrix of size ${M\times M}$ and $\mzer_M$ the \mdim MM zero matrix.
\item $[\va]_m$ and $[\ma]_{m,k}$ respectively denote the $m$th and $(m,k)$ components of $\va$ and $\ma$. Also, $[\ma]_{\cdot,k}$ denotes the $k$th column of $\ma$ and $[\ma]_{\cdot,m:k}$ denotes the matrix form by its columns $m$ to $k$.
\item $\ma^H$ is the Hermitian of $\ma$ and $\ma^\dagger$ its pseudo-inverse.
\item The operator '$\equiv$' introduces new symbols.
\item '$*$' denotes convolution: $(\Fa*\Fb)(t)$ is the convolution of $\Fa(t)$ and $\Fb(t)$. 
\item '$\mathcal{E}\{\cdot\}$' and '$\text{Var}\{\cdot\}$' denote the expectation and variance operators. 
\item $\Fdel(t)$ and $\Fdel_n$ respectively denote Dirac and Kronecker delta. 
\item In the paper, a given ${M\times M}$ matrix $\mpr$ is said to be a projection matrix if $\mpr=\mpr^H$ and $\mpr^2=\mpr$.
\item $\|\mb\|_F$ and $\|\mb\|_2$ denote the Frobenius and 2-norm of a given matrix $\mb$ respectively.
\item $\FO(\cdot)$ is the big-O notation. A function $\Fg(N)$ is $\FO(1/N)$ if there are positive numbers $N_o$ and $A$ such that $|\Fg(N)|\leq A/N$ if $N>N_o$. 
\item $\Fo(\cdot)$ is the little-o notation. A function $\Fg(N)$ is $\Fo(1/N)$ if for any $\epsilon>0$ there is positive $N_o$ such that $|\Fg(N)|\leq \epsilon/N$ if $N>N_o$.
\item In some contexts, '$\cong$' marks an equality that holds if an $\Fo(1/N)$ term is added.
\end{itemize}

The main symbols in the paper are the following matrix functions:

\begin{itemize}
\item $\mr(f_r)$: expected covariance matrix at the $r$th bin, (\ref{eq:308}).
\item $\widehat\mr(f_r)$: sample covariance matrix at $r$th bin, (\ref{eq:387}).
\item $\mpr(f)$: expected signal projection matrix function, (\ref{eq:546}).
\item $\widehat\mpr_0(f_r)$: initial signal projection matrix estimate at $r$th frequency bin, (\ref{eq:10}).
\item $\widehat\mpr_1(f)$: polynomial estimate of signal projection matrix, (\ref{eq:435}).
\item $\widehat\mpr_2(f)$: correction of $\widehat\mpr_1(f)$ to the closest projection matrix, (\ref{eq:328}).
\end{itemize}

\section{Characterization of the signal subspace for wideband subspace estimation, method's outline, and paper organization }
\label{sec:css}

Given a sensor array, the estimation of the subspace spanned by several impinging waves is, in principle, a well-posed problem only in narrow spectral bands, given that the array response varies with frequency. This implies that any extension to wide bands of subspace estimation must take into account the response variation in some way. A relevant feature of this variation is that it is smooth along the band covered by the incoming signals, and the same happens with the subspace spanned by several DOAs.  Let us justify this assertion by performing the following analysis, valid for a generic sensor array.

Consider an array of $M$ sensors and $K$ impinging waves. Also, place the origin of coordinates at the center of the smallest circle or sphere containing all sensors, and let $\tau_{max}$ denote its diameter but measured as a delay at the propagation speed. The response of any of the sensors to any of the DOAs consists of the convolution with a response of the form $a\Fdel(t-\tau)$ for some factor $a$ and delay $\tau$ following ${|\tau|\leq\tau_{max}/2}$. Thus, the spectrum of this response is $a\Fe^{-j2\pi \tau f}$ which is a band-limited signal in the $f$ variable with spectrum inside the interval $[-\tau_{max}/2, \tau_{max}/2]$. This is a smooth function of $f$ and there exist bounds on its derivatives of any order, (Bernstein's inequality \cite[Th. 6.7]{Higgins96}). So, if $\ma(f)$ is the \mdim MK matrix whose $(m,k)$ component is the response of the $m$th sensor to the $k$th DOA, the whole matrix $\ma(f)$ is formed by band-limited functions with spectrum inside $[-\tau_{max}/2, \tau_{max}/2]$, i.e, by smooth functions. In turn, this smoothness of $\ma(f)$ translates into a smooth variation of its span with $f$. To see this point, we must resort to a result in Perturbation Theory, \cite[Ch.2, Th. 1.10]{Kato95}. The span of $\ma(f)$ is uniquely represented by an orthogonal projection matrix $\mpr(f)$, whose components are analytic functions.
\footnote{$\mpr(f)$ is the eigenprojector associated with the $K$ eigenvalues of $\ma(f)\ma(f)^H$. Since this last matrix is Hermitian and all its components are entire functions, we may apply Theorem 1.10 (Chapter 2) in \cite{Kato95} to conclude that $\mpr(f)$ is analytic.}
Note than $\mpr(f)$ is the projection matrix that is usually obtained through the normal equations if $\ma(f)$ has full column rank, i.e, 
\begin{equation}
\label{eq:546}
\mpr(f)=\ma(f)(\ma(f)^H\ma(f))^{-1}\ma(f)^H.
\end{equation}
However, since $\mpr(f)$ is analytic, it is well defined by continuity even at the isolated frequencies at which the column rank of $\ma(f)$ is smaller than $K$ and, additionally, it is approximable in any band $[f_A,f_B]$ by a polynomial of the form 
\begin{equation}
\label{eq:9}
\mpr(f)\approx \sum_{q=0}^Q \mg_qf^q
\end{equation}
where $Q$ is the polynomial order and $\mg_q$ are \mdim MM coefficient matrices (Weiertrass theorem, \cite[Th. 2.4.1]{Phillips03}). Besides, there is an order $Q$ for which the mismatch in (\ref{eq:9}) is negligible for any possible selection of the DOAs. This is so, because the DOAs can be parameterized using a finite set of variables, say angles of arrival $\theta_1$, $\theta_2,\ldots$, $\theta_K$, and the domains of these variables are closed sets (usually finite intervals).  The order $Q$ required depends on the specific array geometry and tolerable mismatch and, in principle, must be determined numerically. We may expect the value of $Q$ to increase with ${f_B-f_A}$, being equal to zero for short bands (narrowband case).  [In Sec. \ref{sec:poq}, we assess the selection of $Q$ for a uniform linear array (ULA) formed by 10 sensors.] 

In order to see the relevance of the smooth variation of $\mpr(f)$, consider a value of $Q$ for which the mismatch of (\ref{eq:9}) is negligible, and recall the usual binning approach in wideband subspace estimation. In this approach, there is a set of sample covariance matrices $\widehat\mr(f_r)$, computed at a set of $R$ frequency bins with central frequencies $f_r$ lying inside $[f_A,f_B]$, \range rR, and each of them
provides a subspace estimate at its corresponding frequency. More precisely, the span of the $K$
eigenvectors associated with the $K$ largest eigenvalues of each $\widehat\mr(f_r)$ approximates the span of $\mpr(f_r)$. But we can describe this approximation in terms of projection matrices. Specifically, if $\widehat\mq_K(f_r)$ is the $M\times K$ matrix formed by these last eigenvectors and following $\widehat\mq_K(f_r)^H\widehat\mq_K(f_r)=\mi_K$, then the matrix
\begin{equation}
\label{eq:536}
\widehat\mpr_0(f_r)\equiv\widehat\mq_K(f_r)\widehat\mq_K(f_r)^H
\end{equation}
approximates $\mpr(f_r)$ component-wise, i.e, $[\widehat\mpr_0(f_r)]_{m,m'}$ is an estimate of $[\mpr(f_r)]_{m,m'}$ for all $m,\,m'=1,\,2,\ldots,\,M$. Now if $\mpr(f)$ is oversampled, i.e, if $R$ is larger than the number of coefficient matrices in (\ref{eq:9}), $Q+1$, then we may estimate $\mg_q$ in (\ref{eq:9}) using a simple procedure, such as least squares. This would give a set of estimates $\widehat\mg_q$ that could then be used to estimate the signal subspace at any frequency in $[f_A,f_B]$, simply by evaluating (\ref{eq:9}) with $\widehat\mg_q$ in place of $\mg_q$.

In the sections that follow, we develop this approach in detail. First, we present the signal model in the next section, including the usual binning approach whose output is a set of projection matrix estimates.  Then, we present the method in Sec. \ref{sec:epm}, which is based on weighted least squares, and show that it is asymptotically consistent. We justify this last fact in Sec. \ref{sec:aft}, where we present asymptotic expressions for the bias and for a bound on the RMS error of the projection matrix function estimate. Finally, we assess the method's quality in Sec. \ref{sec:ne} numerically, both for the estimation of $\mpr(f)$ and for DOA estimation. 

\section{Signal model for wideband subspace estimation}
\label{sec:smw}

Consider the generic scenario, presented in the previous section, consisting of $M$ sensors and $K$ waves impinging from different directions of arrival. Let $\Fs_{bp,k}(t)$ denote the received band-pass signal from the $k$th direction and assume that the receiver's demodulators operate at a frequency $f_o$, so that the lowpass equivalent of $\Fs_{bp,k}(t)$ is
\begin{equation}
\label{eq:339}
\Fs_k(t)\equiv \Fe^{-j2\pi f_o t}\Fs_{bp,k}(t).
\end{equation}
For any array geometry, the effect of the $m$th sensor on $\Fs_k(t)$ can be described through the convolution with an unknown impulse response $\tilde\Fa_{m,k}(t)$; i.e, the signal received at the $m$th sensor is 
\begin{equation}
\label{eq:341}
(\tilde \Fa_{m,k}*\Fs_k)(t).
\end{equation}
$\tilde\Fa_{m,k}(t)$ models models the geometry of the $k$th impinging wave relative to the array and is a time-limited response with support lying inside  $[-\tau_{max}/2,\tau_{max}/2]$ for the delay $\tau_{max}$ discussed in the previous section. Besides, since we may transfer any constant factor from $\tilde\Fa_{m,k}(t)$ to $\Fs_k(t)$ in the convolution in (\ref{eq:341}), we may assume that the Fourier transform of $\tilde\Fa_{m,k}(t)$ is bounded by one; i.e, denoting the Fourier transform of $\tilde\Fa_{m,k}(t)$ by $\Fa_{m,k}(f)$, we have $|\Fa_{m,k}(f)|\leq 1$. Note that this model of the sensor array includes, as a particular case, the usual scenario in which the effect of the $m$th sensor on the $k$th signal is just a delay $\tau_{m,k}\in[-\tau_{max}/2,\tau_{max}/2]$, just by taking
\begin{equation}
\label{eq:382}
\tilde\Fa_{m,k}(t)=\Fe^{-j2\pi f_o \tau_{m,k}}\Fdel(t-\tau_{m,k}).
\end{equation}

Now, if $\Fx_{m}(t)$, \range mM, denotes the lowpass signal from the $m$th sensor, we have by superposition
\begin{equation}
\label{eq:403}
\Fx_m(t)=\sum_{k=1}^{K}(\widetilde\Fa_{m,k}*\Fs_k)(t)+\Feta_m(t),
\end{equation}
where we have included a set of noise processes $\Feta_m(t)$ that we model as circularly-symmetric, complex-white and of zero mean and variance $\sigma^2$.

Next, we proceed to take this last signal model to the frequency domain. For doing this, we assume the following:
\begin{itemize}
\item The signals $\Fs_k(t)$ have significant spectral content inside a band $[f_A,f_B]$.
\item At each sensor, the receiver filters the signals $\Fx_m(t)$ using  $R$ pass-band filters with impulse response $\Fh(t;f_r)$, \range rR, where $f_r$ is the filter's central frequency. The frequencies $f_r$ are distinct, lie in $[f_A,f_B]$, and the filters' pass bands are $[f_r-\Delta f/2,f_r+\Delta f/2]$ for a fixed bandwidth $\Delta f$. Besides, the frequencies form an increasing sequence with spacing at least $\Delta f$, ($f_r\leq f_{r+1}$, \range r{R-1}). For simplicity, we assume unit-energy  responses $\Fh(t,f_r)$, so that the noise power at the output of the corresponding filters have the same power, denoted $\sigma^2$. 

\item $\Delta f$ is selected to ensure the narrowband condition relative to the possible spectra $\Fa_{m,k}(f)$ in the following sense: for any possible DOA and any sensor, it must be ${\Fa_{m,k}(f_1) \approx \Fa_{m,k}(f_2)}$ for any two frequencies following ${|f_2-f_1|\leq\Delta f}$. Since the responses $\Fa_{m,k}(f)$ are time-limited to $[-\tau_{max}/2,\tau_{max}/2]$, from Bernstein's inequality \cite[Th. 11.1.2]{Boas54} we have
%
\begin{equation}
\label{eq:384}
|\Fa'_{m,k}(f)|\leq \pi \tau_{max}.
\end{equation}
Besides this bound is attained by any sensor with delay $\tau_{max}/2$ as can be readily inferred from (\ref{eq:382}). Thus, $\Delta f$ must be much smaller than $1/(\pi \tau_{max})$ to ensure a negligible variation of $\Fa_{m,k}(f)$.

\item The scenario is wideband in the sense that $f_B-f_A>\Delta f$.
\end{itemize}

From (\ref{eq:403}), the output of the $r$th filter at the $m$th sensor  follows the model 
\begin{multline}
\label{eq:365}
(\Fh(\,\cdot\,;f_r)*\Fx_m)(t)\\
=
\sum_{k=1}^{K}(\Fh(\,\cdot\,;f_r)*\widetilde\Fa_{m,k}*\Fs_k)(t)
+(\Fh(\,\cdot\,;f_r)*\Feta_m)(t)
\end{multline}
and, since $\Fa_{m,k}(f_r)$ has a negligible variation in any band of width $\Delta f$, we may approximate the convolution in (\ref{eq:365}) in the following way
\begin{equation} 
\label{eq:307}
(\Fh(\,\cdot\,;f_r)*\widetilde\Fa_{m,k})(t)\approx \Fa_{m,k}(f_r)\Fh(t;f_r),
\end{equation}
and, in turn, write (\ref{eq:365}) as
\begin{multline}
\label{eq:306}
(\Fh(\,\cdot\,;f_r)*\Fx_m)(t)
=\\
\sum_{k=1}^{K}\Fa_{m,k}(f_r)(\Fh(\,\cdot\,;f_r)*\Fs_k)(t)+(\Fh(\,\cdot\,;f_r)*\Feta_m)(t).
\end{multline}

Here is where we can see the subspace structure of $\Fx_m(t)$, given that the signal component on the right-hand side is a linear combination of array response signatures $\va_{m,k}(f_r)$ for any $m$. We may write this model employing the usual notation in narrowband subspace estimation. For this define (\range mM, \range kK),
\begin{align}
  \label{eq:319}
  [\ma(f)]_{m,k}&\equiv \Fa_{m,k}(f),\\\nonumber
  [\vx(t;f_r)]_m&\equiv(\Fh(\,\cdot\,;f_r)*\Fx_m)(t),\\\nonumber
  [\vs(t;f_r)]_k&\equiv(\Fh(\,\cdot\,;f_r)*\Fs_k)(t),\\\nonumber
  [\veps(t;f_r)]_m&\equiv(\Fh(\,\cdot\,;f_r)*\Feta_m)(t).
\end{align}
Then, (\ref{eq:306}) can be written as
\begin{equation}
  \label{eq:386}
  \vx(t;f_r)= \ma(f_r)\vs(t;f_r)+\veps(t;f_r), r=1,\,2,\ldots,\,R.
\end{equation}
This is a set of $R$ independent narrowband models that allow for the computation of subspace estimates \cite[Sec. 9.3]{VanTreesP4}. Specifically, for each output $\vx(t;f_r)$, we have the following processing:
\begin{enumerate}
\item If we model the components of $\vs(t;f_r)$ as wide-sense stationary processes which are independent of $\veps(t;f_r)$, then the expected covariance of $\vx(t,f_r)$, $\mathcal E\{\vx(t;f_r) \vx(t;f_r)^H\}$, is given by
\begin{equation}
\label{eq:308}
\mr(f_r)\equiv\ma(f_r)\mr_{s}(f_r)\ma(f_r)^H+\sigma^2\mi_M,
\end{equation}
where  $[\mr_{s}(f_r)]_{k,k'}$ is the cross-power density spectrum of components $k$ and $k'$ of $\vs(t;f_r)$, (${k,k'=1,}\,2,\ldots\,K$).

\item The ML estimator of $\mr(f_r)$ can be readily computed from $N$ samples of $\vx(t;f_r)$ and is given by
\begin{equation}
\label{eq:387}
\widehat\mr(f_r)\equiv \frac{1}{N}\sum_{n=0}^{N-1}\vx(nT;f_r)\vx(nT;f_r)^H.
\end{equation}
\item Let $\widehat\mq_K(f_r)$ denote the \mdim MK matrix formed by the $K$ eigenvectors associated with the $K$ largest eigenvalues of $\widehat\mr(f_r)$. The span of $\widehat\mq_K(f_r)$ is the ML estimate of the corresponding span of $\mr(f_r)$ which, in turn, is the span of $\ma(f_r)$ if $\mr_s(f_r)$ is non-singular. 
\end{enumerate}  

$\widehat\mq_K(f_r)$ provides an estimate of $\mpr(f_r)$ which is 
\begin{equation}
  \label{eq:10}
  \widehat\mpr_0(f_r)\equiv \widehat\mq_K(f_r)\widehat\mq_K(f_r)^H.
\end{equation}

The set of matrices $\widehat\mpr_0(f_r)$, \range rR, is the data from which we proceed to estimate the whole function $\mpr(f)$ in $[f_A,f_B]$ in the next section.

\section{Estimation of the projection matrix function}
\label{sec:epm}

Let us assume that an order $Q$ has been selected such that there exist polynomials of the form in (\ref{eq:9}) with negligible mismatch and that ${Q<R}$. (\ref{eq:9}) and the narrowband estimates $\per$ in (\ref{eq:10}) allow us to pose a linear regression model for an arbitrary component $(m,m')$ of $\widehat\mpr_0(f_r)$, 
\begin{equation}
\label{eq:423}
[\widehat\mpr_0(f_r)]_{m,m'}\approx\sum_{q=0}^Q[\mg_q]_{m,m'}f_r^q,\;\;r=1,\,2\ldots,\,R,
\end{equation}
in which $[\mg_q]_{m,m'}$, \rangect qQ, is the set of unknown parameters. The estimates $[\per]_{m,m'}$ are independent given that $f_{r+1}-f_r\geq\Delta f$, \range r{R-1}. Besides, we prove in the next section that $\widehat\mpr_0(f_r)$ is a consistent estimator of $\mpr(f_r)$, given that 
\begin{align}
\label{eq:8}
 \mathcal E\{\widehat\mpr_0(f_r)\}&=\mpr(f_r)+\FO(1/N)\\
 \nonumber \text{Var}\{\widehat\mpr_0(f_r)\}&=\FO(1/N),
\end{align}
where the variance is taken component-wise. All this suggests to pose a weighted least-squares cost function for the estimation of the coefficients $[\mg_q]_{m,m'}$ with weights $w_r$, i.e, the function
\begin{equation}
\label{eq:424}
\sum_{r=1}^R w_r^2
\Big|[\widehat\mpr_0(f_r)]_{m,m'}-\sum_{q=0}^Q[\widehat\mg_q]_{m,m'}f_r^q\Big|^2,
\end{equation}
[For a possible set of coefficients $w_r$, see (\ref{eq:550}).] Its minimizer is the estimate
\begin{equation}
\label{eq:425}
[\widehat\mg_q]_{m,m'}\equiv\sum_{r=1}^R[\mb]_{q,r}[\widehat\mpr_0(f_r)]_{m,m'},
\end{equation}
where (\range {r,r'}R; $q=0,\,1,\ldots,\,Q$)
\begin{align}
\label{eq:426}
  \mb&\equiv (\mv^H\mlam^2\mv)^{-1}\mv^H\mlam^2,\;\;[\mv]_{r,q+1}\equiv f_r^q,\\ [\mlam]_{r,r'}&\equiv
   w_r\Fdel_{r-r'}.
\end{align}
Since $\mb$ is independent of $m$ and $m'$, we may write (\ref{eq:425}) as
\begin{equation}
\label{eq:427}
\widehat\mg_q\equiv\sum_{r=1}^R[\mb]_{q,r}\widehat\mpr_0(f_r)
\end{equation}
and the estimate of $\mpr(f)$ at any $f$ is
\begin{equation}
\label{eq:428}
\widehat\mpr_1(f) \equiv\sum_{q=0}^Q \widehat\mg_qf^q.
\end{equation}
$\widehat\mpr_1(f)$ can be written in terms of $\per$ by substituting (\ref{eq:427}) into (\ref{eq:428}). We have 
\begin{equation}
\label{eq:435}
\widehat\mpr_1(f) \equiv\sum_{r=1}^R\Fd_r(f)\widehat\mpr_0(f_r)
\end{equation}
where 
\begin{equation}
\label{eq:436}
\Fd_r(f)=\sum_{q=0}^Q [\mb]_{q,r}f^q.
\end{equation}

Let us now analyze the quality of the estimate $\widehat\mpr_1(f)$ by considering its mean and variance. Since the mismatch of the polynomial approximation is negligible, ${R>Q}$ and the weights $w_r$ are positive, we have that (\ref{eq:435}) holds for $\mpr(f)$,
\begin{equation}
\label{eq:540}
\mpr(f)=\sum_{r=1}^R\Fd_r(f)\mpr(f_r).
\end{equation}
So, from (\ref{eq:8}), (\ref{eq:435}) and this last equation, we have
\begin{align}
\label{eq:541}
\ex{\widehat\mpr_1(f)}&=\mpr(f)+\FO(1/N)\\
\text{Var}\{\widehat\mpr_1(f)\}&=\FO(1/N).
\end{align}
Thus, $\widehat\mpr_1(f)$ is a consistent estimate of $\mpr(f)$. 

For any $f$ in $\fab$, $\widehat\mpr_1(f)$ is a projection matrix only approximately and it may be necessary to provide an exact projection matrix in some applications. This drawback can be solved by taking the rank-$K$ projection matrix lying closest to $\widehat\mpr_1(f)$ as the final signal projection matrix estimate at each frequency; i.e, the final signal projection matrix estimate is
\begin{equation}
  \label{eq:328}
  \widehat\mpr_2(f)\equiv
  \arg\min_{\textrm{
     $\genfrac{}{}{0pt}{0}{\textrm{rank}\;K}{\textrm{proj. matrix}\;\mpr}$
      }}\|\widehat\mpr_1(f)-\mpr\|^2_F.
\end{equation}
It can be easily checked that $\widehat\mpr_2(f)$ is just the signal projection matrix of $\widehat\mpr_1(f)$, [i.e, the projection matrix associated with the $K$ largest eigenvalues of $\widehat\mpr_1(f)$]. Though the computation of $\widehat\mpr_2(f)$ from $\widehat\mpr_1(f)$ requires an additional eigenvalue decomposition, we may expect that the number of such decompositions is small in practice, given that $\widehat\mpr(f)$ can be well approximated in $[f_A,f_B]$ by a $Q$-order polynomial. We assess this point in Secs. \ref{sec:nem} and \ref{sec:pi} numerically. 

\section{Assessment of the first two moments of $\widehat\mpr_0(f_r)$}
\label{sec:aft}
In this section, we present asympotic expressions for the mean and for a bound on the variance of $\widehat\mpr_0(f_r)$. These expressions prove (\ref{eq:8}) as a corollary and provide a possible set of coefficients $w_r$ in (\ref{eq:424}). For simplicity, let us suppress the dependency on $f_r$ in writing in the rest of this section, i.e, let us write $\mr$ rather than $\mr(f_r)$ and $\widehat\mr$ rather than $\widehat\mr(f_r)$, etc. Let us start by recalling the existing results on the perturbation of the eigenvalues of a sample covariance matrix \cite{Brillinger01,Kaveh86}. For this, write the eigenvalue decomposition of $\mr$ as
\begin{equation}
\label{eq:490}
\mr=\sum_{m=1}^M\lambda_m\vq_m\vq_{m}^H,
\end{equation}
where $\lambda_m$ is the $m$th eigenvalue in decreasing order and $\vq_m$ is a corresponding eigenvector, $\vq_m^H \vq_{m'} = \Fdel_{m-m'}$, ($m,\,m'=1,\ldots,\,M$). Due to the model in (\ref{eq:308}), we have ${\lambda_m=\sigma^2}$, $m=K+1,,\ldots,\, M$. For the sample covariance matrix $\widehat\mr$ in (\ref{eq:387}), this same decomposition takes the form 
\begin{equation}
\label{eq:491}
\widehat\mr=\sum_{m=1}^M\widehat\lambda_m\widehat\vq_m\widehat\vq_{m}^H,
\end{equation}
where ``$\widehat{\;\;}$'' marks the estimates of the corresponding parameters. Let $\widehat\veps_m$ denote the estimation error for $\vq_m$, i.e, 
\begin{equation}
\label{eq:1}
\widehat\vq_m=\vq_m+\widehat\veps_m.
\end{equation}

In order to recall the asymptotic expressions for the first two moments of $\widehat\veps_m$,
define first the coefficients
\begin{equation}
\label{eq:5}
b_{m,\ell}\equiv
\begin{cases}
  \dfrac{\lambda_m\lambda_\ell}{N (\lambda_m-\lambda_\ell)^2}& \text{if $m\neq \ell$}\\
  0& \text{otherwise}
  \end{cases}
\end{equation}
From Eqs. (12) to (14) in \cite{Kaveh86}, we have the following asymptotic expressions, where ``$\cong$'' denotes $\Fo(1/N)$ equalities (\range{m,\,m'}{M}):
\begin{equation}
\label{eq:2}
   \mathcal E\{\widehat\veps_m\}
  \cong\Big(-\frac{1}{2}
\sum_{\ell=1}^Mb_{m,\ell}\Big)\vq_m,
\end{equation}
\begin{equation}
\label{eq:3}
\mathcal E\{\widehat\veps_m
                  \widehat\veps_{m'}^H\}
  \cong\Fdel_{m-m'}\sum_{\ell=1}^Mb_{m,\ell}\vq_\ell\vq_\ell^H,
\end{equation}
\begin{equation}
\label{eq:4}
\mathcal E\{\widehat\veps_m
\widehat\veps_{m'}^T\} \cong-
(1-\Fdel_{m-m'})b_{m,m'}\vq_{m'}\vq_m^T.
\end{equation}

In App. \ref{sec:fso} , we use these expressions to prove the formula 
\begin{equation}
\label{eq:7}
\mathcal E\{\widehat\mpr_0\}=\mpr+\mq\mlam_b\mq^H+\Fo(1/N),
\end{equation}
where ${[\mq]_{\cdot,m}\equiv \vq_m}$, (\range mM) and $\mlam_b$ is the diagonal matrix
\begin{equation}
\label{eq:6}
  [\mlam_b]_{m,m'}\equiv 
  \begin{cases}
    \D{-\Fdel_{m-m'}\sum_{m=K+1}^Mb_{m,\ell}&\text{if $m\leq K$}}\\
    \D{{}\;\;\,\Fdel_{m-m'}\sum_{k=1}^Kb_{k,\ell}&\text{if $K<m\leq M$}}.
  \end{cases}
\end{equation}

Since $b_{k,\ell}$ is $\FO(1/N)$, (\ref{eq:7}) implies that  $\widehat\mpr_0$ is asymptotically unbiased provided there is some separation between the signal and noise subspaces, i.e, ${\lambda_K-\lambda_{K+1}>0}$. To see this last point, note that all coefficients $b_{k,\ell}$ in (\ref{eq:6}) depend on one signal eigenvalue and one noise eigenvalue and, therefore, we have
\begin{equation}
\label{eq:493}
b_{k,\ell}=\frac{\lambda_k\lambda_\ell}{N(\lambda_k-\lambda_\ell)^2}\leq \frac{\lambda_1\lambda_{K+1}}{N(\lambda_K-\lambda_{K+1})^2}.
\end{equation}

The variance of the components of $\widehat\mpr_0$ can be computed using the results in \cite{Brillinger01,Kaveh86} through a laborious derivation and the result seems to depend on $\mpr$. However, we may proceed in an indirect way by bounding  $\|\widehat\mpr_0-\mpr\|_F^2$, given that 
%
\begin{equation}
\label{eq:504}
\max_{m,m'}\big|[\widehat\mpr_0-\mpr]_{m,m'}\big|^2\leq \|\widehat\mpr_0-\mpr\|_F^2.
\end{equation}
In App. \ref{ap:dbe}, we prove the following asymptotic bound
\begin{multline}
\label{eq:430}
\mathcal E \{\|\widehat\mpr_0-\mpr\|_F^2\}\\
\leq \frac{8}{N(\lambda_K-\lambda_{K+1})^2}\sum_{k=1}^K\lambda_k
\sum_{\ell=1}^{M-K}\lambda_{K+\ell}+\Fo\Big(\frac{1}{N}\Big).
\end{multline}
Recalling that ${\lambda_m=\sigma^2}$ if ${K<m\leq M}$, we may combine the last two inequalities to obtain a bound on the quadratic error for any component $(m,m')$, 
\begin{equation}
\label{eq:505}
\mathcal  E \Big\{\big|[\widehat\mpr_0-\mpr]_{m,m'}\big|^2\Big\}
\leq\frac{8(M-K)\sigma^2}{N(\lambda_K-\sigma^2)^2}\sum_{k=1}^K\lambda_k+\Fo\Big(\frac{1}{N}\Big).
\end{equation}

This inequality provides a possible choice of coefficients $w_r$ in (\ref{eq:424}), if we eliminate the factors independent of either the eigenvalues $\lambda_k$ or $\sigma^2$. Specifically, we have the possible coefficients
\begin{equation}
\label{eq:549}
\widehat w_r\equiv \frac{\widehat\lambda_K-\widehat\sigma^2}{\widehat\sigma}\Bigg(\sum_{k=1}^K\widehat\lambda_k\Bigg)^{-1/2},
\end{equation}
where $\widehat\sigma$ is a deviation estimate obtained from the covariance matrix $\widehat\mr$ of one or more frequency bins. For a single frequency bin, this estimate would be
\begin{equation}
\label{eq:550}
\widehat\sigma\equiv \Bigg(\frac{1}{M-K}\sum_{m=K+1}^M \widehat\lambda_m\Bigg)^{1/2}.
\end{equation}

Since, from (\ref{eq:7}), the mean of $[\widehat\mpr_0-\mpr]_{m,m'}$ is $\FO(1/N)$, (\ref{eq:505}) is also valid for the variance, i.e,
\begin{equation}
\label{eq:537}
\text{Var}\{[\widehat\mpr_0-\mpr]_{m,m'}\}
\leq\frac{8(M-K)\sigma^2}{N(\lambda_K-\sigma^2)^2}\sum_{k=1}^K\lambda_k
+\Fo\Big(\frac{1}{N}\Big).
\end{equation}

The bound in (\ref{eq:430}) has an interesting interpretation. Since ${\lambda_K-\sigma^2}$ is the power gap between the signal and noise subspaces, the factor
\begin{equation}
\label{eq:432}
\frac{1}{\lambda_K-\sigma^2}\sum_{k=1}^K\lambda_k
\end{equation}
is the signal power measured in number of gaps and
\begin{equation}
\label{eq:433}
\frac{(M-K)\sigma^2}{\lambda_K-\sigma^2}
\end{equation}
is the noise power again measured in number of gaps. Therefore, the bound in (\ref{eq:537}) is proportional to the product of these two relative powers and decreases as $1/N$.

\section{Application of the proposed method to wideband DOA estimation}
\label{sec:apm}

In this section, we apply the method in Sec. \ref{sec:epm} to wideband DOA estimation in a  ULA. Let us first particularize the signal model in Sec. \ref{sec:smw} to this type of array, then recall two wideband DOA estimators, IC-MUSIC and MTOPS, and finally adapt these estimators to the method proposed in this paper.

In a ULA, the time-domain response of the $m$th sensor to the $k$th DOA in (\ref{eq:341}) is
\begin{equation}
\label{eq:402}
\tilde \Fa_{k,m}(t)=\Fe^{-j2\pi f_o \tau_m\gamma_k}\Fdel(t-\tau_m\gamma_k)
\end{equation}
where
\begin{itemize}
\item $f_o$ is the array's central frequency, so that the sensor spacing is $c/(2f_o)$ where $c$ is the propagation speed,
\item $\gamma_k\equiv\sin(\theta_k)$ and $\theta_k$ is the angle of arrival relative to the broadside,
\item  $\tau_m$ is the delay associated with the $m$th sensor along the array. If  $\tau_{max}$ denotes the array diameter, measured as a delay, then 
\begin{equation}
\label{eq:543}
\tau_m\equiv-\frac{\tau_{max}}{2}+(m-1)\frac{\tau_{max}}{M-1},\,m=1,\,2,\ldots,\,M.
\end{equation}
\end{itemize}
The array response $\ma(f)$ in (\ref{eq:319}) is just the Fourier transform of (\ref{eq:402}), (\range mM, \range kK),

\begin{equation}
  \label{eq:202}
[\ma(f,\gamma)]_{m,k}\equiv \Fe^{-j2\pi (f_o+f)\tau_m\gamma_k},
\end{equation}
where
\begin{equation}
\label{eq:205}
[\vgam]_k\equiv \gamma_k,\; \textrm{\range kK},
\end{equation}
and where we have written $\ma(f,\vgam)$ rather than $\ma(f)$ to show the dependency on the parameters $\gamma_k$ explicitly.
Finally, we may write the model in (\ref{eq:308}) for $\mr(f_r)$ as
\begin{equation}
\label{eq:204}
\mr(f_r)=\ma(f_r,\vgam)\mr_s(f_r)\ma(f_r,\vgam)^H+\sigma^2\mi_M.
\end{equation}
The problem of estimating the angles of arrival $\theta_k$ can now be cast as the problem of estimating the parameters $\gamma_k$, given that there is a one-to-one relationship between $\theta_k$ and $\gamma_k$, $\gamma_k=\sin(\theta_k)$.

In this paper, we consider the estimation of the DOA parameters $\gamma_k$ by means of the following two methods:

\begin{itemize}
\item {\bf IC-MUSIC, \cite[Sec. 4.4.3]{Tuncer09}.} In this estimator, the $K$ DOA estimates are given by the abscissa of the main $K$ local maxima of the pseudo-spectrum
\begin{equation}
\label{eq:185}
\Fphi(\gamma)\equiv \sum_{r=1}^R\|\per\va(f_r,\gamma)\|^2,
\end{equation}
where
\begin{equation}
\label{eq:548}
[\va(f,\gamma)]_m\equiv \Fe^{-j2\pi (f_o+f)\tau_m\gamma},\,m=1,\,2,\ldots,\,M.
\end{equation}

\item {\bf MTOPS, \cite{Shaw16}.} This estimator starts by computing a set of $M\times K$ matrices $\widehat\mur(f_r)$ and another set of $M\times(M-K)$ matrices $\widehat\mv(f_r)$, \range rR, whose columns are ortho-normal bases of the signal and noise subspaces respectively; i.e, the columns of $\widehat\mur(f_r)$ span the subspace associated with $\per$ and

\begin{equation}
\label{eq:186}
[\widehat\mur(f_r),\widehat\mv(f_r)]^H[\widehat\mur(f_r),\widehat\mv(f_r)]=\mi_M.
\end{equation}

Note that we may view the matrix pair  $(\widehat\mur(f_r), \widehat\mv(f_r))$ as a function of  $\per$, given that one such pair can be easily computed from $\widehat\mpr_0(f_r)$. The MTOPS estimator uses the pseudo-spectrum

\begin{equation}
\label{eq:188}
\mu(\gamma)\equiv \Big\|\big[\me_2(\gamma),\me_3(\gamma),
\ldots,\,\me_R(\gamma)\big]\Big\|^2_F,
\end{equation}
where ($r=2,\,3,\ldots, R$)
\begin{align}
  \label{eq:189}
\me_r(\gamma)&\equiv\widehat\mur(f_1)^H\mathrm{diag}(\va(f_r-f_1,\gamma))\widehat\mv(f_r).
\end{align}

The MTOPS DOA estimates are the smallest $K$ local minima of $\mu(\gamma)$. 
\end{itemize}

These estimators can be easily adapted to the method proposed in this paper, simply by applying them to a set of samples  of the projection function estimate  $\widehat\mpr_2(f)$; i.e, the matrices $\widehat\mpr_0(f_r)$, (\range rR), in IC-MUSIC and MTOPS would be replaced by a set of matrices $\widehat\mpr_2(f'_{r})$, (\range {r}{R'}), where the frequencies $f'_r$ are regularly spaced in $[f_A,f_B]$ and $R'$ is close to $Q+1$. Note that the computation of the matrices $\widehat\mpr_2(f'_{r})$ involves $R'$ eigenvalue decompositions. However, this additional complexity is small if $R\gg Q$, given that we may expect $R'$ to be close to $Q$. 

\section{Numerical examples}
\label{sec:ne}

We have carried out several numerical examples for a 10-sensor ULA following the model in Sec. \ref{sec:apm}. The sub-sections that follow contain the main results:

\begin{itemize}
\item Sec. \ref{sec:mpn} contains a list of the main parameters in the numerical examples.
\item In Sec. \ref{sec:poq}, we assess the selection of $Q$ using Chebyshev interpolation in order to upper-bound the mismatch of the polynomial approximation in (\ref{eq:9}) for several values of $Q$.
\item In Sec. \ref{sec:nep}, we evaluate the approximation error of one component of $\mpr(f)$ using the corresponding component of $\widehat\mpr_1(f)$.
\item In Sec. \ref{sec:rae}, we evaluate the same error but for the whole matrix $\widehat \mpr_1(f)$ using the error measure in (\ref{eq:198}) and in the RMS sense. 
\item Finally, in Secs. \ref{sec:nem} and \ref{sec:pi}, we combine the proposed method with IC-MUSIC and MTOPS for DOA estimation.
\end{itemize}

\subsection{Main parameters in the numerical examples}
\label{sec:mpn}

 The parameters in the numerical examples were the following:
\vspace{0.5em}

\noindent {\bf Sensor array.} Linear array of ${M=10}$ sensors with half wavelength spacing.

\noindent {\bf Central frequency.} $f_o= 2.4$ GHz.

\noindent {\bf Maximum delay along the array.}
\begin{displaymath}
  \tau_{max}=\frac{M-1}{2f_o}=1.875\; \text{nsec}.
\end{displaymath}

\noindent {\bf DOA parameters.} The DOA was parameterized in terms of $\gamma$ rather than $\theta$, where $\gamma=\sin(\theta)$, following the approach in Sec. \ref{sec:apm}.

\noindent {\bf Received signals.} Linearly-modulated signals of the form

\begin{equation}
  \label{eq:173}
  \sum_{n=-\infty}^\infty a_n\Fg(t-n T_{ch})
\end{equation}

where

\begin{itemize}
\item $a_n$ are zero-mean, independent complex Gaussian noise samples of variance equal to 1.
\item $\Fg(t)$ is a raised-cosine pulse with chip period $T_{ch}\equiv 2.6$ nsec and roll-off factor $\beta\equiv 0.25$.
\end{itemize}

\noindent {\bf Sampling period.} $T\equiv T_{ch}/2$.

\noindent {\bf Signals' bandwidth relative to $f_o$.} The signals' two-sided bandwidth $B$ followed ${B/f_o=0.2}$, i.e, ${B=0.48}$ GHz. However, in the numerical examples, only the band in which $\Fg(t)$ has flat spectrum was used, i.e, the band $[-B_1/2,B_1/2]$, where ${B_1\equiv}{(1-\beta)/T_{ch}}{=0.288}$ GHz. So, the relative bandwidth employed was ${B_1/f_o}{=0.12}$.

\noindent {\bf Number of slots.} $N_{sl}=50$.

\noindent {\bf Number of samples per slot.} $N=1024$.

\noindent {\bf Number of frequency bins.} The number of frequency bins was fixed to ${R=41}$ and they were equally spaced in $[-B_1/2,B_1/2]$. 

\noindent {\bf Directions of arrival (DOAs).} Two cases have been assessed:

\begin{itemize}
\item Three DOAs given by

\begin{equation}
\label{eq:197}
\vgam=[0.1,0.27,0.82]^T.
\end{equation}

\item Two DOAs of the form

\begin{equation}
\label{eq:206}
\vgam=[0.1,0.1+\Delta\gamma]^T,
\end{equation}

where the increment $\Delta\gamma$ is a simulation parameter. 
\end{itemize}

\noindent {\bf Signal-to-noise ratio (SNR).} The SNRs in the numerical examples are equal to the total signal energy at frequency $f_o$ divided by the corresponding noise energy. 

\noindent {\bf Least squares weights $w_r$.} Equal to 1. 

\noindent {\bf Generation method for numerical trials.} The numerical trials were generated using the frequency domain model in (\ref{eq:386}) for reducing the simulation time. However, the results were validated by generating these trials in the time domain and checking that the results are consistent if $N\rightarrow \infty$. 

\noindent {\bf Number of Monte Carlo trials.} 2200.

\subsection{Polynomial order $Q$ versus mismatch for a ten-sensor uniform linear array}
\label{sec:poq}

We have computed an upper bound for the approximation error in (\ref{eq:9}) by means of   Chebyshev interpolation applied to $\mpr(f)$ in range ${f\in[f_A,f_B]}$ with ${Q+1}$ nodes,  where the array response matrix is given by (\ref{eq:202}), \cite[Ch. 6]{Mason02}. 
This interpolation scheme delivers a polynomial approximation in the $f$ variable of the form
\begin{equation}
\label{eq:502}
\mpr(f)\approx \sum_{q=0}^Q \mg_q(Q,\gamma_1,\ldots,\gamma_K)f^q
\end{equation}
for $f$ in $[f_A,f_B]$ and fixed $Q$, $\gamma_1$, $\ldots$, $\gamma_K$. Specifically, we have evaluated the error in (\ref{eq:502}) for any value of the variables involved.  Table \ref{tab:1} shows the maximum error in (\ref{eq:502}) for several values of $Q$ and $K$, i.e the error measure
\begin{equation}
  \label{eq:13}
\max\limits_{m,m',f,\gamma_1,\ldots,\gamma_K}\Big|[\mpr(f)-\sum_{q=0}^Q \mg_q(Q,\gamma_1,\ldots,\gamma_K)f^q]_{m,m'}\Big|. 
\end{equation}
\begin{table}[t]
  \centering
  \begin{tabular}[c]{|c|l|l|l|l|}
    \hline
    \diagbox[innerleftsep=1mm,width=1cm,innerwidth=0.5cm,
    innerleftsep=1mm,innerrightsep=1mm]{$Q$}{\raisebox{-0.5mm}{$K$}} & \multicolumn{1}{c|}{1}&
\multicolumn{1}{c|}{2}&
\multicolumn{1}{c|}{3}&
                        \multicolumn{1}{c|}{4}\\    \hline
\rule{0pt}{1em}0& 0.1486 & 0.2624 & 0.3313 & 0.3766 \\
\rule{0pt}{1em}1& 0.06228 & 0.09993 & 0.1106 & 0.114 \\
\rule{0pt}{1em}2& 0.01739 & 0.02558 & 0.03154 & 0.03407 \\
\rule{0pt}{1em}3& 0.003732 & 0.007473 & 0.01038 & 0.01325 \\
\rule{0pt}{1em}4& 0.0006299 & 0.00211 & 0.003046 & 0.00411 \\
\rule{0pt}{1em}5& 0.00008891 & 0.0007461 & 0.001155 & 0.00163 \\
    \hline
  \end{tabular}
  \vspace{1em}
  \caption{Chebyshev interpolation error using measure in (\ref{eq:13}).}
  \label{tab:1}
\end{table}
Note that, since the components of $\mpr(f)$ are bounded by one, this interpolation scheme provides significant accuracy even for small values of $Q$. Table \ref{tab:2} shows the error measure
\begin{equation}
  \label{eq:14}
\sup\limits_{f,\gamma_1,\ldots,\gamma_K}  \|\mpr(f)-\sum_{q=0}^Q \mg_q(Q,\gamma_1,\ldots,\gamma_K)f^q\|_F
\end{equation}
where, again, the supremum is taken over all variables involved. This is the maximum error that would be incurred if $\mpr(f)$ were replaced by its polynomial approximation in the computation of $\mpr(f)\vx$ for unit-norm vectors $\vx$. The conclusion is the same. 

\begin{table}[t]
  \centering
  \begin{tabular}[c]{|c|l|l|l|l|}
    \hline
    \diagbox[innerleftsep=1mm,width=1cm,innerwidth=0.5cm,
    innerleftsep=1mm,innerrightsep=1mm]{$Q$}{\raisebox{-0.5mm}{$K$}} & \multicolumn{1}{c|}{1}&
\multicolumn{1}{c|}{2}&
\multicolumn{1}{c|}{3}&
                        \multicolumn{1}{c|}{4}\\    \hline
\rule{0pt}{1em}0& 0.5051 & 0.6319 & 0.6578 & 0.6794 \\
\rule{0pt}{1em}1& 0.1552 & 0.1986 & 0.2247 & 0.2454 \\
\rule{0pt}{1em}2& 0.03203 & 0.0652 & 0.07654 & 0.08669 \\
\rule{0pt}{1em}3& 0.005908 & 0.02142 & 0.02551 & 0.03059 \\
\rule{0pt}{1em}4& 0.0008885 & 0.006993 & 0.00854 & 0.01067 \\
\rule{0pt}{1em}5& 0.0001158 & 0.00227 & 0.002881 & 0.003813 \\
\hline
  \end{tabular}
  \vspace{1em}
  \caption{Chebyshev interpolation error using measure in (\ref{eq:14}).}
  \label{tab:2}
\end{table}

\subsection{Approximation of $\widehat\mpr_1(f)$ and $\widehat\mpr_2(f)$ to $\mpr(f)$}
\label{sec:nep}

In this section, we assess qualitatively the error in approximating the true projection matrix $\mpr(f)$ using either $\widehat\mpr_0(f)$, $\widehat\mpr_1(f)$ or $\widehat\mpr_2(f)$ for a single component of these matrices and the DOAs in (\ref{eq:197}). The objective of this assessment is to show  the improvement that can be achieved graphically. 
\begin{figure}
\centering{\includegraphics{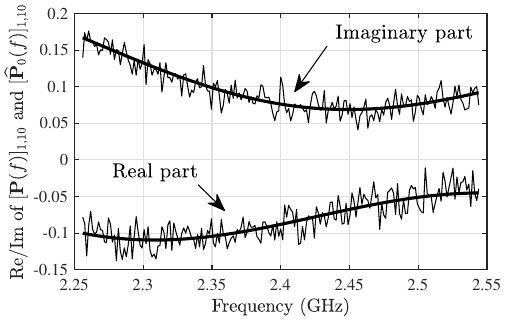}}
\caption{\label{fig:6} Real and imaginary part of component $[\mpr(f)]_{1,10}$ (thick lines) and its associated noisy estimates from $[\hat\mpr_0(f)]_{1,10}$ (thin lines).}
\end{figure}
Fig. \ref{fig:6} shows  the real and imaginary parts of component $[\mpr(f)]_{1,10}$ and the state-of-the-art estimate $[\widehat\mpr_0(f)]_{1,10}$ in one realization of the numerical example for ${\textrm{SNR}=8~\textrm{dB}}$.
The smooth thick curves are the components of $[\mpr(f)]_{1,10}$ and the noisy thin curves the corresponding components of $[\widehat\mpr_0(f)]_{1,10}$. Note that $[\widehat\mpr_0(f)]_{1,10}$ approximates $[\mpr(f)]_{1,10}$ but with some error, fundamentally due to the variation of the received signals' sample spectra.  Though this figure only shows one component of $\mpr(f)$ and $\widehat\mpr_0(f)$,  the trends  also hold for the rest of components, i.e, the whole matrix $\mpr(f)$ is a smooth function of $f$ and  ${\mpr(f)\approx\widehat\mpr_0(f)}$.

\begin{figure}
\centering{\includegraphics{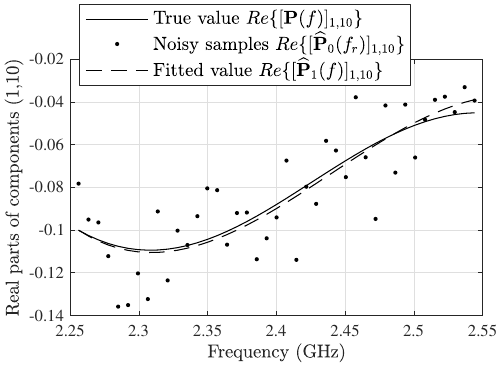}}
\caption{\label{fig:7} Result of fitting a  polynomial of order ${Q=3}$ to ${R=41}$ equally-spaced estimates $\mathrm{Re}\{[\widehat\mpr_0(f_r)]_{1,10}\}$. }
\end{figure}
Fig. \ref{fig:7} shows the result of fitting a polynomial of order ${Q=3}$ to the real part of ${R=41}$ equally-spaced samples of $\mathrm{Re} \{[\widehat\mpr_0(f_r)]_{1,10}\}$. The continuous curve is the true value $[\pxw]_{1,10}$ and the dashed curve the fitted value $[\widehat\mpr_1(f)]_{1,10}$. Note that the fitted value is a significantly better estimate of $[\mpr(f)]_{1,10}$ along the frequency band than the initial estimates (dots).  

\begin{figure}
\centering{\includegraphics{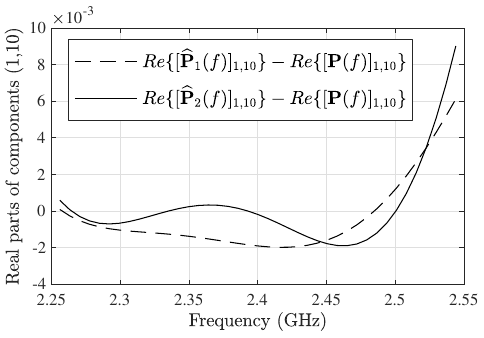}}
\caption{\label{fig:10} Error in approximating $\textrm{Re}\{[\mpr(f)]_{1,10}\}$ using either $\textrm{Re}\{[\widehat\mpr_1(f)]_{1,10}\}$ or $\textrm{Re}\{[\widehat\mpr_2(f)]_{1,10}\}$.}
\end{figure}
The final correction in (\ref{eq:328}) for obtaining $\widehat\mpr_2(f)$ from $\widehat\mpr_1(f)$ produces a slight variation, that can be readily seen in Fig. \ref{fig:10} for the real part of component $(1,10)$. This figure shows the error in approximating $\mpr(f)$ using either $\widehat\mpr_1(f)$ or $\widehat\mpr_2(f)$ for component $(1,10)$. Note that the curve is smooth for $\widehat\mpr_1(f)$ and $\widehat\mpr_2(f)$ and that the approximation error is small in both cases. 

\subsection{RMS approximation error of $\widehat\mpr_1(f)$ versus the number of sample covariance matrices $R$}
\label{sec:rae}
\begin{figure}
  \centering{\includegraphics{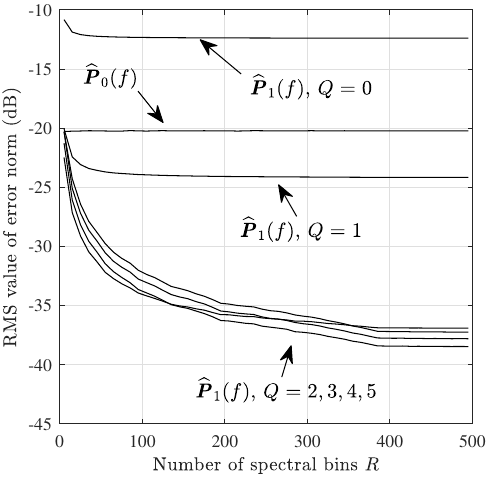}}
\caption{\label{fig:1} RMS value of error norm in (\ref{eq:198}) versus the number of spectral bins $R$. }
\end{figure}
Fig. \ref{fig:1} shows the approximation error for the whole projection matrix in the example of the previous sub-section, where the error norm is
\begin{equation}
\label{eq:198}
\Big(\frac{1}{R}\sum_{r=1}^R\|\mpr(f_r)-\widehat\mpr_1(f_r)\|_2^2\Big)^{1/2}.
\end{equation}
Note that, except for $Q=0$, $\widehat\mpr_1(f)$ outperforms $\widehat\mpr_0(f)$ by a significant margin that can reach $17~\textrm{dB}$ for a high number of bins $R$. 
\begin{figure}
\centering{\includegraphics{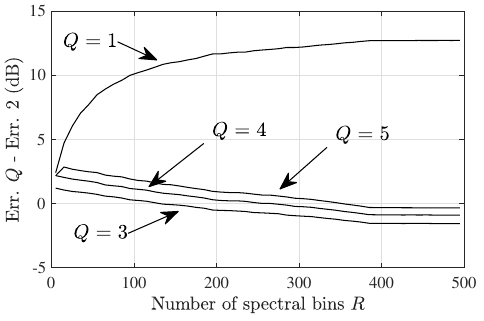}}
\caption{\label{fig:2} Difference in dBs between the RMS error of $\widehat\mpr_1(f)$ for $Q=2$ and the same error for $Q\neq 2$ versus the number of covariance matrices $R$. }
\end{figure}
Fig. \ref{fig:2} shows the curves in Fig. \ref{fig:1} for ${Q\neq 2}$ minus the curve for ${Q=2}$ in dBs. This figure allows us to see what value of $Q$ performs best versus the number of bins $R$. As can be readily seen, ${Q=1}$ is the best choice up to ${R=10}$  (value below zero in ``${Q=1}$'' curve), while ${Q=2}$ is the best choice between $R=11$ and ${R=250}$, and ${Q=3}$ is the best choice for ${R>250}$.   

\subsection{Improvement in DOA separation}
\label{sec:nem}

In order to assess the effect of the proposed method on the separation of DOA estimates, we have evaluated the RMS error in the estimation of $\gamma_1$ for varying $\Delta\gamma$ in (\ref{eq:206}) using several variants of IC-MUSIC and MTOPS for ${\textrm{SNR}= -10~\textrm{dB}}$. The variants were the following: 
\begin{itemize}
\item {\bf Standard IC-MUSIC}. Standard IC-MUSIC estimator using the pseudo-spectrum in (\ref{eq:185}). 

\item {\bf 41-bin IC-MUSIC}. Standard IC-MUSIC estimator but with projection matrices $\widehat\mpr_0(f_r)$ replaced with their estimates $\widehat\mpr_2(f_r)$.
\item {\bf 5-bin IC-MUSIC}. IC-MUSIC estimator applied to $R'=5$ projection matrices $\widehat\mpr_2(f'_r)$ as proposed at the end of Sec. \ref{sec:apm}. The frequencies $f'_{r}$ formed a regular grid covering the band $[f_o-B_1/2,f_o+B_1/2]$ and the IC-MUSIC pseudo-spectrum was (\ref{eq:185}) but with frequencies $f'_r$, \range r{R'}.

\item {\bf 1-bin IC-MUSIC}. The same estimator but computed from the single projection matrix $\widehat\mpr_2(f_o)$, i.e, with $f'_1=f_o$ and $R'=1$. 
\end{itemize}

And the MTOPS estimators were:
\begin{itemize}
\item {\bf Standard MTOPS}. MTOPS estimator computed from the pseudo-spectrum in (\ref{eq:188}). Since this pseudo-spectrum has numerous local peaks, the peak lying closest to 1-bin IC-MUSIC was selected as estimate. (See \cite{Shaw16} for comments on this drawback of MTOPS.) 

\item {\bf 41-bin MTOPS}. Standard MTOPS estimator but with projection matrices $\widehat\mpr_0(f_r)$ replaced with their estimates $\widehat\mpr_2(f_r)$.
\item {\bf 5-bin MTOPS}. MTOPS estimator applied to ${R'=5}$ projection matrices $\widehat\mpr_2(f'_r)$, (Sec. \ref{sec:apm}), where the frequencies $f'_{r}$ formed a regular grid covering the band $[f_o-B_1/2,f_o+B_1/2]$.
  
\end{itemize}

\begin{figure}
  \subfigure[\label{fig:3a} $R'=41$ projection matrices $\widehat\mpr_2(f'_r)$.]{ 
    \includegraphics{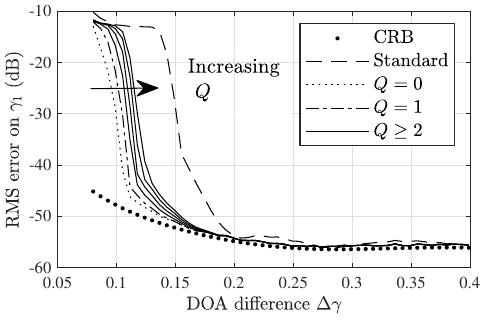}}

    \subfigure[\label{fig:3b} $R'=5$ projection matrices $\widehat\mpr_2(f'_r)$.]{ 
      \includegraphics{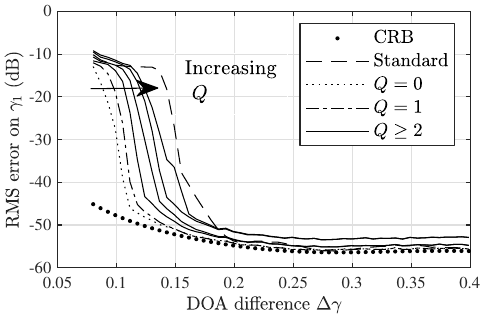}}

     \subfigure[\label{fig:3c} $R'=1$ projection matrices $\widehat\mpr_2(f'_r)$.]{\includegraphics{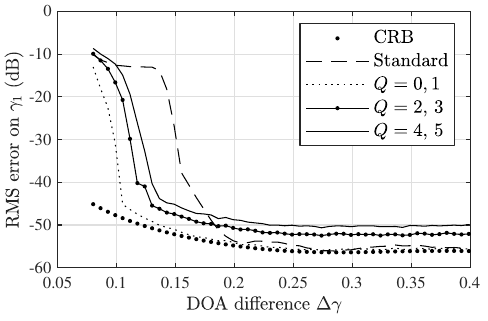}}
\caption{\label{fig:3} RMS DOA error of IC-MUSIC in the estimation of $\gamma_1$ for the DOAs in (\ref{eq:206}) versus the DOA difference $\Delta\gamma$. }
\end{figure}

Fig \ref{fig:3a} shows the performance of 41-bin IC-MUSIC versus that of standard IC-MUSIC.  Note that the proposed method improves the smallest $\Delta \gamma$ for which the two DOAs are separable. With the standard estimator, we have $\Delta \gamma=0.15$, while the adapted method with $Q=1$ provides $\Delta \gamma=0.11$ roughly. Larger values of $Q$ imply a larger threshold for $\Delta\gamma$. Fig. \ref{fig:3b} shows the same comparison for 5-bin IC-MUSIC and the behavior is the same, but with a more noticeable degradation with increasing $Q$.

\begin{figure}
  \subfigure[\label{fig:15a} $R'=41$ projection matrices $\widehat\mpr_2(f'_r)$.]{ 
    \includegraphics{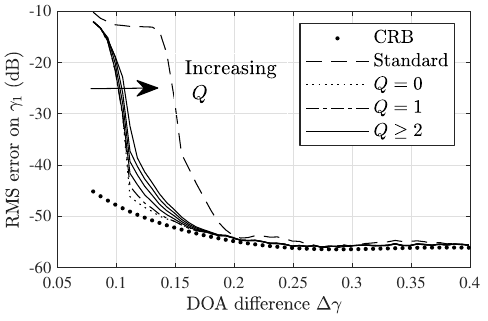}}

    \subfigure[\label{fig:15b} $R'=5$ projection matrices $\widehat\mpr_2(f'_r)$.]{ 
      \includegraphics{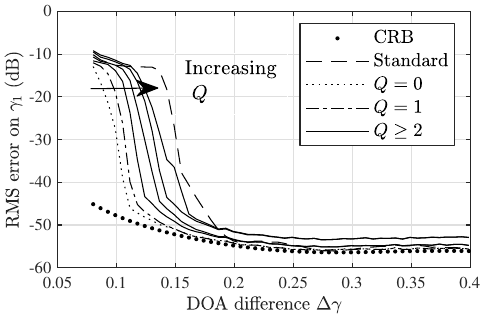}}

\caption{\label{fig:3} RMS DOA error of MTOPS in the estimation of $\gamma_1$ for the DOAs in (\ref{eq:206}) versus the DOA difference $\Delta\gamma$. }
\end{figure}
Figs. \ref{fig:3c}, \ref{fig:15a} and \ref{fig:15b} show the same comparison for 5-bin IC-MUSIC, 41-bin MTOPS and 5-bin MTOPS, respectively, and the conclusions that can be drawn are similar. 

\subsection{Performance improvement in SNR threshold}
\label{sec:pi}

\begin{figure}
  \subfigure[\label{fig:17a} $R'=41$ projection matrices $\widehat\mpr_2(f'_r)$.]{ 
    \includegraphics{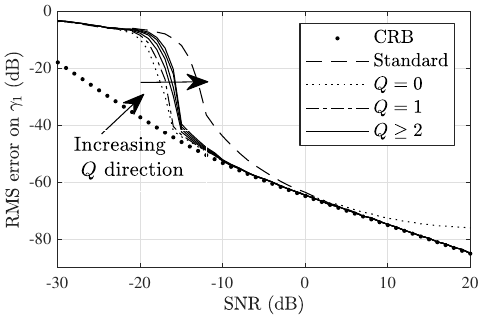}}

    \subfigure[\label{fig:17b} $R'=5$ projection matrices $\widehat\mpr_2(f'_r)$.]{ 
      \includegraphics{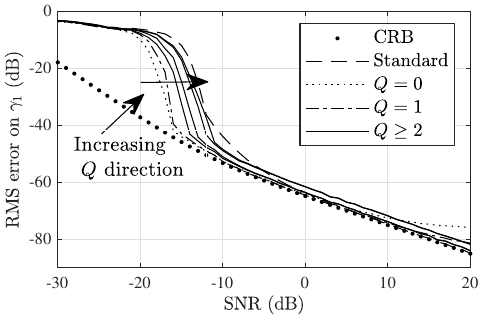}}

     \subfigure[\label{fig:17c} $R'=1$ projection matrices $\widehat\mpr_2(f'_r)$.]{\includegraphics{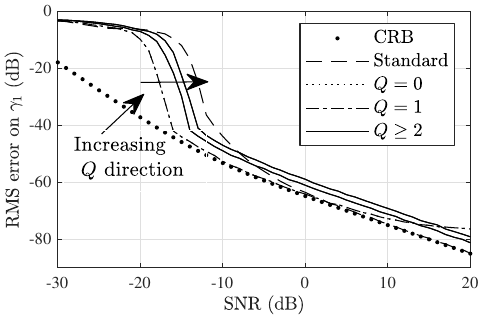}}
\caption{\label{fig:3} RMS DOA error of IC-MUSIC in the estimation of $\gamma_1$ for the DOAs in (\ref{eq:197}) versus the SNR. }
\end{figure}

\begin{figure}
  \subfigure[\label{fig:20a} $R'=41$ projection matrices $\widehat\mpr_2(f'_r)$.]{ 
    \includegraphics{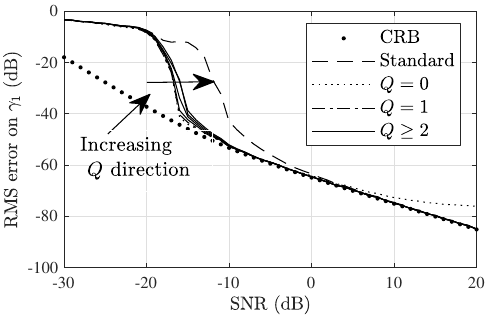}}

    \subfigure[\label{fig:20b} $R'=5$ projection matrices $\widehat\mpr_2(f'_r)$.]{ 
      \includegraphics{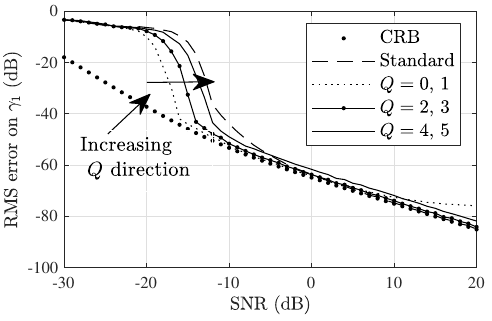}}

\caption{\label{fig:3} RMS DOA error of MTOPS in the estimation of $\gamma_1$ for the DOAs in (\ref{eq:197}) versus the SNR. }
\end{figure}

We have repeated the numerical example in the previous sub-section, but for the three DOAs in (\ref{eq:197}), and assessed the RMS error of $\gamma_1$ for varying SNR. Fig. \ref{fig:17a} shows the performance of 41-bin IC-MUSIC. Note that the behavior is similar to the one in the previous sub-section, i.e, the method provides an improvement of roughly 7 dBs in the SNR threshold. Additionally, we can see that the mismatch for the polynomial approximation can be preceived in the high RMS error for high SNRs and $Q=0$. Figs \ref{fig:17b}, \ref{fig:17c}, \ref{fig:20a}, \ref{fig:20b} show the result of the same assessment for the remaining estimators in the previous section, 5-bin and 1-bin IC-MUSIC, and 41-bin and 5-bin MTOPS, respectively, with similar conclusions.

\section{Conclusions}

We have presented a method for characterizing the signal subspace through a polynomial approximation of the signal projection matrix, valid in a given frequency band. Fundamentally, the method consists of fitting a polynomial to a set of sample signal projection matrices, obtained through the usual binning approach in wideband subspace estimation. The resulting polynomial provides an approximate signal projection matrix at any frequency in the band, which can then be used to improve the quality of wideband DOA estimators such as IC-MUSIC and MTOPS. We have presented asymptotic bounds for the bias and RMS error of the polynomial estimate and we have assessed its performance in several numerical examples. 

\appendices

\section{First- and second-order moments of sample covariance matrix eigenvectors}
\label{sec:fso}

Let us first compute the first- and second-order moments of $\widehat \vq_m$. From (\ref{eq:2}), we have
\begin{equation}
\label{eq:451}
   \mathcal E\{\widehat\vq_m\}
  \cong \vq_m+\Big(-\frac{1}{2}
\sum_{\ell=1}^Mb_{m,\ell}\Big)\vq_m=c_m\vq_m,
\end{equation}
where we define the coefficients
\begin{equation}
\label{eq:461}
c_m\equiv 1-\frac{1}{2}
\sum_{\ell=1}^Mb_{m,\ell}.
\end{equation}
From (\ref{eq:3}) and (\ref{eq:451}), we have
\begin{align}
\label{eq:453}
\mathcal E\{&\widehat\vq_m\widehat\vq_{m'}^H\}=\mathcal E\{(\vq_m+\widehat\veps_m)(\vq_{m'}+\widehat\veps_{m'})^H\} \nonumber \\
            &=\vq_m\vq_{m'}^H+\vq_mE\{\widehat\veps_{m'}^H\}+E\{\widehat\veps_m\}\vq_{m'}^H
              +E\{\widehat\veps_m\widehat\veps_{m'}^H\} \nonumber\\
            &=(1+c_m+c_{m'})\vq_m\vq_{m'}^H+
              \Fdel_{m-m'}\sum_{\ell=1}^Mb_{m,\ell}\vq_\ell\vq_\ell^H.
\end{align}
If $m=m'$ this expression reduces to
\begin{equation}
\label{eq:535}
\mathcal E\{\widehat\vq_m\widehat\vq_m^H\}=
(1+2c_m)\vq_m\vq_{m}^H+\sum_{\ell=1}^Mb_{m,\ell}\vq_\ell\vq_\ell^H.
\end{equation}

Adding up (\ref{eq:535}) for \range mK, we obtain a formula for $\mathcal E\{\widehat \mpr_0\}$, noting that $\mpr=\sum_{k=1}^K\vq_k\vq_k^H$:
\begin{align}
\label{eq:454}
\mathcal E\{\widehat \mpr_0\}&\cong \sum_{k=1}^K \Bigg(
(1+2c_k)\vq_k\vq_k^H+\sum_{\ell=1}^Mb_{k,\ell}\vq_\ell\vq_\ell^H
                             \Bigg) \nonumber \\
  &=\sum_{k=1}^K \Bigg(
    \Big(1-\sum_{\ell=1}^Mb_{k,\ell}\Big)\vq_k\vq_k^H+\sum_{\ell=1}^Mb_{k,\ell}\vq_\ell\vq_\ell^H
    \Bigg) \nonumber \\
  &=\mpr-\sum_{k=1}^K 
   \sum_{\ell=1}^Mb_{k,\ell}\vq_k\vq_k^H+\sum_{k=1}^K \sum_{\ell=1}^Mb_{k,\ell}\vq_\ell\vq_\ell^H.
\end{align}
And doing the same for $m=K+1,\,\ldots,\,M$, we obtain
\begin{multline}
  \label{eq:480}
  \mathcal E\{\mi_{M}-\widehat \mpr_0\}\cong
  \mi_M-\mpr \\ -\sum_{m=K+1}^M 
   \sum_{\ell=1}^Mb_{m,\ell}\vq_m\vq_m^H+\sum_{m=K+1}^M \sum_{\ell=1}^Mb_{m,\ell}\vq_\ell\vq_\ell^H
\end{multline}

The last two formulas involve coefficients $b_{k,\ell}$ in which both $k$ and $\ell$ are either smaller than ${K+1}$ or larger than $K$, i.e, coefficients computed from pairs of eigenvalues associated to either the signal or noise subspace. Such coefficients $b_{k,\ell}$ can be arbitrarily large,  given that the only condition on the eigenvalues $\lambda_m$ is that there is a significant gap $\lambda_{K}- \lambda_{K+1}$ between the signal and noise eigenvalues. We solve this drawback by writing $\mathcal E\{\widehat\mpr_0\}$ in terms of $\mathcal E\{(\mi_M-\mpr)\widehat \mpr_0\}$ and $\mathcal E\{\mpr(\mi_M-\widehat\mpr_0)\}$. We have from (\ref{eq:454}),
\begin{equation}
\label{eq:457}
\mathcal E\{(\mi_M-\mpr)\widehat \mpr_0\}\cong \sum_{k=1}^K \sum_{\ell=K+1}^Mb_{k,\ell}\vq_\ell\vq_\ell^H
\end{equation}
and from (\ref{eq:480}),
\begin{equation}
\label{eq:458}
\mathcal E\{\mpr(\mi_M-\widehat \mpr_0)\}\cong
\sum_{m=K+1}^M \sum_{\ell=1}^Kb_{m,\ell}\vq_\ell\vq_\ell^H.
\end{equation}
Finally, combining the last two equations, we have
\begin{multline}
\label{eq:459}
\mathcal E\{\widehat\mpr_0\}
=\mathcal E\{\mpr+(\mi_M-\mpr)\widehat \mpr_0-\mpr(\mi_M-\widehat \mpr_0)\}\\
                          =\mpr+\mathcal E\{(\mi_M-\mpr)\widehat \mpr_0\}-\mathcal E\{\mpr(\mi_M-\widehat \mpr_0)\}\\
                          \cong \mpr+ \sum_{\ell=K+1}^M\Big(\sum_{k=1}^Kb_{k,\ell}\Big)\vq_\ell\vq_\ell^H
                          \\
                          -\sum_{\ell=1}^K \Big(\sum_{m=K+1}^Mb_{m,\ell}\Big)\vq_\ell\vq_\ell^H.
\end{multline}
The matrix form of this expression is (\ref{eq:7}).

\section{Derivation of bound on expected quadratical error}
\label{ap:dbe}

The eigenvalue decomposition of $\mr$ in (\ref{eq:490}) can be written as
\begin{equation}
\label{eq:513}
\mr=\mq\mlam\mq^H
\end{equation}
where $\mq$ is unitary, $[\mq]_m\equiv\vq_m$, and $\mlam$ is a diagonal matrix with components $[\mlam]_{m,m'}=\Fdel_{m-m'}\lambda_m$, \range {m,m'}M. The sample covariance matrix $\widehat\mr$ can be described as the average of $N$  independent, complex normal \mdim M1 vectors $\vx_n$, of zero mean and covariance $\mr$,
\begin{equation}
\label{eq:514}
\widehat\mr=\frac{1}{N}\sum_{n=1}^N \vx_n\vx_n^H.
\end{equation}

Next, consider the vectors $\vs_n\equiv\mq^H\vx_n$ which are also independent, complex normal and of zero mean, but of covariance matrix $\mlam$. The sample covariance matrix of these vectors is
\begin{equation}
\label{eq:515}
\widehat\mc\equiv \frac{1}{N}\sum_{n=1}^N \vs_n\vs_n^H.
\end{equation}
Given a realization $\widehat\mr$ and its corresponding $\widehat\mc$, let us bound the error in approximating $\mpr$ using $\widehat\mpr_0$ by resorting to a perturbation theory result in \cite{Stewart90}. First, define the following measure for the dissimilarity between the spans of $\mpr$ and $\widehat\mpr_0$:

\begin{equation}
\label{eq:12}
\widehat\eta\equiv\frac{\|\widehat\mc_{sn}\|_F^2}{(\lambda_{K}-\lambda_{K+1})^2},
\end{equation}
where $\widehat\mc_{sn}$ is the block formed by the intersection of the first $K$ rows  and last $M-K$ columns of $\widehat\mc$, i.e, $\widehat\mc_{sn} \equiv [\widehat\mc]_{1:K,K+1:M}$.
Combining theorems 2.1 and 3.1 of  \cite{Stewart90}, we have that if $\widehat \eta<1/4$ then 
\begin{equation}
\label{eq:517}
\|\widehat\mpr_0-\mpr\|_F^2\leq 8\,\widehat\eta.
\end{equation}
(See also comments on page 232 of \cite{Stewart90}.)

In order to turn (\ref{eq:517}) into an asymptotic inequality, let us first compute the first two moments of the components of $\widehat\mc_{sn}$ and then the mean and variance of $\widehat\eta$. For simplicity, let $c_{k,\ell}$ and $s_{n,k}$ denote $[\widehat\mc]_{k,\ell}$ and $[\vs_n]_{k}$ respectively. For any indices $k$, $\ell$, $p$ and $q$, lying between 1 and $M$ and following $k\neq \ell$ and $p\neq q$, we have:

\begin{itemize}
\item $\mathcal E\{c_{k,\ell}\}=0$ given that $k\neq \ell$ and
\begin{equation}
\label{eq:519}
\mathcal E\{c_{k,\ell}\}=[\ex{\widehat\mc}]_{k,\ell}=[\mlam]_{k,\ell}=0.
\end{equation}
\item The formula
\begin{equation}
\label{eq:520}
\ex{c_{k,\ell}c_{p,q}}=\frac{1}{N}\Fdel_{k-q}\Fdel_{\ell-p}\lambda_k\lambda_\ell.
\end{equation}
To prove this result, recall the formula for the expectation of the product of four complex normal random variables $a_k$, \cite{Janssen88}:
\begin{multline}
\label{eq:521}
\ex{a_1a_2a_3a_4}=\ex{a_1a_2}\ex{a_3a_4}+\ex{a_1a_3}\ex{a_2a_4}\\
+\ex{a_1a_4}\ex{a_2a_3}-\ex{a_1}\ex{a_2}\ex{a_3}\ex{a_4}.
\end{multline}
The proof is the following. We have
\begin{multline}
   \label{eq:15}
  \hspace{-1.cm}\ex{c_{k,\ell}c_{p,q}}=\mathcal E\Big\{ [\frac{1}{N}\sum_{n=1}^N \vs_n\vs_n^H]_{k,\ell} [\frac{1}{N}\sum_{n'=1}^N \vs_{n'}\vs_{n'}^H]_{p,q}\Big\} \\
  \hspace{-1.cm}= \frac{1}{N^2}\sum_{n=1}^N\sum_{n'=1}^N\ex{ s_{n,k}s_{n,\ell}^*s_{n',p}s_{n',q}^*}
  \\ \hspace{-1cm} =\frac{1}{N^2}\sum_{n=1}^N\sum_{n'=1}^N\Big( \ex{s_{n,k}s_{n,\ell}^*}\ex{s_{n',p}s_{n',q}^*} \\
  \hspace{-1cm}+\ex{ s_{n,k}s_{n',p}}\ex{ s_{n,\ell}^*s_{n',q}^*}  +\ex{ s_{n,k}s_{n',q}^*}\ex{s_{n,\ell}^*s_{n',p}} \\
 \hspace{-1cm} -\ex{s_{n,k}}\ex{s_{n,\ell}}\ex{s_{n',p}}\ex{s_{n',q}}\Big)
\end{multline}
In this parenthesis, we have:
\begin{itemize}
\item The first term is zero because $\ex{s_{n,k}s_{n,\ell}^*}=[\mlam]_{k,\ell}=0$.
\item If $n\neq n'$ the second term is zero because $\vs_n$ and $\vs_{n'}$ are independent and $\ex{\vs_n}=0$. If $n=n'$ this term is zero because $\ex{\vs_n\vs_n^T}=\vzer_M$.
\item The third term is zero if ${n\neq n'}$ because $\vs_n$ is independent of $\vs_{n'}$ and $\ex{\vs_n}=0$. And, if $n=n'$, then it is also zero if $k\neq q$ or $\ell\neq p$, because $\ex{\vs_n\vs_n^H}=\mlam$ is a diagonal matrix. Thus, we have that the third term is equal to
\begin{multline}
  \Fdel_{n-n'}\Fdel_{k-q} \Fdel_{\ell-p}\ex{s_{n,k}s_{n,k}^*}\ex{s_{n,\ell}s_{n,\ell}^*}\\
  =
  \Fdel_{n-n'}\Fdel_{k-q} \Fdel_{\ell-p}\lambda_k\lambda_\ell.
\end{multline}
\item The fourth term is zero because $\ex{\vs_n}=0$.
\end{itemize}
So, in summary, we have
\begin{multline}
  \label{eq:16} \ex{c_{k,\ell}c_{p,q}}=\frac{1}{N^2}\sum_{n=1}^N\sum_{n'=1}^N\Fdel_{n-n'}\Fdel_{k-q} \Fdel_{\ell-p}\lambda_k\lambda_\ell\\
  =\frac{1}{N}\Fdel_{k-q} \Fdel_{\ell-p}\lambda_k\lambda_\ell.
\end{multline}

\item We also have
\begin{equation}
\label{eq:523}
\ex{c_{k,\ell}c_{p,q}^*}=\frac{1}{N}\Fdel_{k-p}\Fdel_{\ell-q}\lambda_k\lambda_\ell.
\end{equation}

The proof is the following. We have
\begin{multline}
  \hspace{-1.cm}\ex{c_{k,\ell}c^*_{p,q}}=\mathcal E\Big\{ [\frac{1}{N}\sum_{n=1}^N \vs_n\vs_n^H]_{k,\ell} [\frac{1}{N}\sum_{n'=1}^N \vs_{n'}^*\vs_{n'}^T]_{p,q}\Big\}\nonumber \\
  \hspace{-1.cm}= \frac{1}{N^2}\sum_{n=1}^N\sum_{n'=1}^N\ex{ s_{n,k}s_{n,\ell}^*s_{n',p}^*s_{n',q}} 
  \\ \hspace{-1cm}
=\frac{1}{N^2}\sum_{n=1}^N\sum_{n'=1}^N\ex{ s_{n,k}s_{n,\ell}^*s_{n',q}s_{n',p}^*} 
  \nonumber
\end{multline}
Comparing this expression with the second line of (\ref{eq:15}), we can readily see that
\begin{equation}
\label{eq:526}
\ex{c_{k,\ell}c^*_{p,q}}=\ex{c_{k,\ell}c_{q,p}}.
\end{equation}
Therefore, from (\ref{eq:16}), we obtain (\ref{eq:523}).

\end{itemize}

Let us now compute the mean of $\widehat\eta$. From (\ref{eq:12}) and (\ref{eq:523}), we have
\begin{align}
\label{eq:528}
\ex{\widehat\eta}&=\frac{1}{(\lambda_K-\lambda_{K+1})^2}\sum_{k=1}^{K}\sum_{\ell=K+1}^{M}\ex{c_{k,\ell}c_{k,\ell}^*} \nonumber \\
&=\frac{1}{N(\lambda_K-\lambda_{K+1})^2}\sum_{k=1}^{K}\sum_{\ell=K+1}^{M}
\lambda_k\lambda_{\ell}.
\end{align}
Regarding the second-order moment, it follows the formula:
\begin{align}
  \ex{\widehat\eta^2}&=\frac{1}{(\lambda_K-\lambda_{K+1})^4} \nonumber \\
  &\cdot\sum_{k=1}^{K}\sum_{\ell=K+1}^{M} \sum_{k'=1}^{K}\sum_{\ell'=K+1}^{M} \ex{c_{k,\ell}c_{k,\ell}^*c_{k',\ell'}^*c_{k',\ell'}}. 
\end{align}
Expanding the summand using (\ref{eq:521}), we have 
\begin{multline}
\label{eq:529}
\ex{c_{k,\ell}c_{k,\ell}^*c_{k',\ell'}^*c_{k',\ell'}}=
\ex{c_{k,\ell}c_{k,\ell}^*}\ex{c_{k',\ell'}^*c_{k',\ell'}}\\ +
\ex{c_{k,\ell}c_{k',\ell'}^*}\ex{c_{k,\ell}^*c_{k',\ell'}}+
\ex{c_{k,\ell}c_{k',\ell'}}\ex{c_{k,\ell}^*c_{k',\ell'}^*}.
\end{multline}
Note that this is a sum of products consisting of factors of the form in either (\ref{eq:520}) or (\ref{eq:523}). As a consequence, all these products are $\FO(1/N^2)$ and we have 
\begin{equation}
\label{eq:530}
\ex{\widehat\eta^2}=\FO(1/N^2).
\end{equation}

Finally, let us derive the asymptotic inequality. Start by decomposing the expectation of $\|\widehat\mpr_0-\mpr \|_F^2$ by conditioning on $\widehat\eta^2<1/4$:
\begin{align}
  \label{eq:531}
  \mathcal E \{\|&\widehat\mpr_0-\mpr\|_F^2\}=\nonumber \\
  \mathcal E &\Big\{\|\widehat\mpr_0-\mpr\|_F^2\;|\;\widehat\eta^2\leq 1/4\Big\}\Big(1-\mathcal{P}(\widehat\eta^2> 1/4)\Big)\nonumber \\
  &+\mathcal E \Big\{\|\widehat\mpr_0-\mpr\|_F^2\;|\;\widehat\eta^2>1/4\Big\}\mathcal{P}(\widehat\eta^2> 1/4)  \\
  = \mathcal E& \Big\{\|\widehat\mpr_0-\mpr\|_F^2\;|\;\widehat\eta^2\leq 1/4\Big\}
  \nonumber \\
  &+\Big(\mathcal E \Big\{\|\widehat\mpr_0-\mpr\|_F^2\;|\;\widehat\eta^2> 1/4\Big\}
  \nonumber \\ &\hspace{1em}-\mathcal E \Big\{\|\widehat\mpr_0-\mpr\|_F^2\;|\;\widehat\eta^2\leq 1/4\Big\}\Big)
  \cdot\mathcal{P}(\widehat\eta^2> 1/4). \nonumber
\end{align}
From (\ref{eq:517}), we have that the first term follows
\begin{equation}
\label{eq:532}
\mathcal E \{\|\widehat\mpr_0-\mpr\|_F^2\;|\;\widehat\eta^2\leq 1/4\}
\leq 8\,\ex{\widehat\eta}.
\end{equation}
Regarding the second term, the expectations inside the parenthesis are bounded, because projection matrices have components bounded by one. Besides, we may apply Markov's inequality and use (\ref{eq:530}) to obtain
\begin{equation}
\label{eq:533}
\mathcal{P}(\widehat\eta> 1/4)=\mathcal{P}(\widehat\eta^2> 1/16)< 16\ex{\widehat\eta^2}=\FO(1/N^2).
\end{equation}
So, we have that the whole second term in (\ref{eq:531}) is $\FO(1/N^2)$ and, therefore, is $\Fo(1/N)$. In summary, recalling (\ref{eq:528}), we obtain the asymptotic inequality
\begin{align}
\label{eq:534}
  \mathcal E &\{\|\widehat\mpr_0-\mpr\|_F^2\} \nonumber \\
  &\leq \frac{8}{N(\lambda_K-\lambda_{K+1})^2}\sum_{k=1}^{K}\sum_{\ell=K+1}^{M}
\lambda_k\lambda_{\ell}+\Fo\Big(\frac{1}{N}\Big).
\end{align}

\bibliographystyle{IEEEbib}

\bibliography{../../../Utilities/LaTeX/Bibliography}

\begin{thebibliography}{10}

\bibitem{VanTreesP4}
Harry~L. van Trees,
\newblock {\em Detection, Estimation, and Modulation Theory. Part {I}{V},
  Optimum array processing},
\newblock John Wiley \& Sons, Inc, first edition, 2002.

\bibitem{Wang85}
H.~Wang and M.~Kaveh,
\newblock ``Coherent signal-subspace processing for the detection and
  estimation of angles of arrival of multiple wide-band sources,''
\newblock {\em IEEE Transactions on Acoustics, Speech, and Signal Processing},
  vol. 33, no. 4, pp. 823--831, Aug 1985.

\bibitem{Valaee95}
S.~Valaee and P.~Kabal,
\newblock ``Wideband array processing using a two-sided correlation
  transformation,''
\newblock {\em IEEE Transactions on Signal Processing}, vol. 43, no. 1, pp.
  160--172, Jan 1995.

\bibitem{Yasar08}
T.~K. Yasar and T.~E. Tuncer,
\newblock ``Wideband {D}{O}{A} estimation for nonuniform linear arrays with
  {W}iener array interpolation,''
\newblock in {\em 2008 5th IEEE Sensor Array and Multichannel Signal Processing
  Workshop}, July 2008, pp. 207--211.

\bibitem{Zeng10}
W.~J. Zeng and X.~L. Li,
\newblock ``High-resolution multiple wideband and nonstationary source
  localization with unknown number of sources,''
\newblock {\em IEEE Transactions on Signal Processing}, vol. 58, no. 6, pp.
  3125--3136, June 2010.

\bibitem{Wax84}
M.~Wax, Tie-Jun Shan, and T.~Kailath,
\newblock ``Spatio-temporal spectral analysis by eigenstructure methods,''
\newblock {\em IEEE Transactions on Acoustics, Speech, and Signal Processing},
  vol. 32, no. 4, pp. 817--827, Aug 1984.

\bibitem{Doron93}
M.~A. Doron, A.~J. Weiss, and H.~Messer,
\newblock ``Maximum-likelihood direction finding of wide-band sources,''
\newblock {\em IEEE Transactions on Signal Processing}, vol. 41, no. 1, pp.
  411--414, Jan 1993.

\bibitem{Yip02}
Lean Yip, Joe~C. Chen, Ralph~E. Hudson, and Kung Yao,
\newblock ``Cramer-{R}ao bound analysis of wideband source localization and
  {D}{O}{A} estimation,''
\newblock in {\em International Symposium on Optical Science and Technology}.
  International Society for Optics and Photonics, 2002, pp. 304--316.

\bibitem{Chen02}
Joe~C. Chen, Ralph~E. Hudson, and Kung Yao,
\newblock ``Maximum-likelihood source localization and unknown sensor location
  estimation for wideband signals in the near-field,''
\newblock {\em IEEE Transactions on Signal Processing}, vol. 50, no. 8, pp.
  1843--1854, 2002.

\bibitem{Yip08}
L.~Yip, C.~E. Chen, R.~E. Hudson, and K.~Yao,
\newblock ``{D}{O}{A} estimation method for wideband color signals based on
  least-squares joint approximate diagonalization,''
\newblock {\em Proceedings of Sensor Array and Multichannel Signal Processing},
  pp. 104--107, 2008.

\bibitem{Selva18b}
J.~Selva,
\newblock ``Efficient wideband {D}{O}{A} estimation through function evaluation
  techniques,''
\newblock {\em IEEE Transactions on Signal Processing}, vol. 66, no. 12, pp.
  3112--3123, June 2018.

\bibitem{McWirther07}
J.~G. {McWhirter}, P.~D. {Baxter}, T.~{Cooper}, S.~{Redif}, and J.~{Foster},
\newblock ``An {E}{V}{D} {A}lgorithm for {P}ara-{H}ermitian {P}olynomial
  {M}atrices,''
\newblock {\em IEEE Transactions on Signal Processing}, vol. 55, no. 5, pp.
  2158--2169, May 2007.

\bibitem{Alrmah11}
M.~A. {Alrmah}, S.~{Weiss}, and S.~{Lambotharan},
\newblock ``An extension of the {M}{U}{S}{I}{C} algorithm to broadband
  scenarios using a polynomial eigenvalue decomposition,''
\newblock in {\em 2011 19th European Signal Processing Conference}, Aug 2011,
  pp. 629--633.

\bibitem{Weiss13}
S.~{Weiss}, M.~{Alrmah}, S.~{Lambotharan}, J.~G. {McWhirter}, and M.~{Kaveh},
\newblock ``Broadband angle of arrival estimation methods in a polynomial
  matrix decomposition framework,''
\newblock in {\em 2013 5th IEEE International Workshop on Computational
  Advances in Multi-Sensor Adaptive Processing (CAMSAP)}, Dec 2013, pp.
  109--112.

\bibitem{Redif17}
Soydan Redif, Stephan Weiss, and John~G. McWhirter,
\newblock ``Relevance of polynomial matrix decompositions to broadband blind
  signal separation,''
\newblock {\em Signal Processing}, vol. 134, pp. 76 -- 86, 2017.

\bibitem{Boufounos11}
Petros~T. Boufounos, Paris Smaragdis, and Bhiksha Raj,
\newblock ``Joint sparsity models for wideband array processing,''
\newblock {\em Proceedings SPIE}, vol. 8138, pp. 1--10, 2011.

\bibitem{Shen14}
Q.~Shen, W.~Liu, W.~Cui, S.~Wu, Y.~D. Zhang, and M.~G. Amin,
\newblock ``Group sparsity based wideband {D}{O}{A} estimation for co-prime
  arrays,''
\newblock in {\em 2014 IEEE China Summit International Conference on Signal and
  Information Processing (ChinaSIP)}, July 2014, pp. 252--256.

\bibitem{Shen15}
Q.~Shen, W.~Liu, W.~Cui, S.~Wu, Y.~D. Zhang, and M.~G. Amin,
\newblock ``Low-complexity direction-of-arrival estimation based on wideband
  co-prime arrays,''
\newblock {\em IEEE/ACM Transactions on Audio, Speech, and Language
  Processing}, vol. 23, no. 9, pp. 1445--1456, Sept 2015.

\bibitem{Higgins96}
J.~R. Higgins,
\newblock {\em Sampling Theory in {F}ourier and signal analysis.
  {F}oundations.},
\newblock Oxford Science Publications, first edition, 1996.

\bibitem{Kato95}
Tosio Kato,
\newblock {\em Perturbation theory for linear operators},
\newblock Springer, 1995.

\bibitem{Phillips03}
G.~M. Phillips,
\newblock {\em Interpolation and approximation by polynomials},
\newblock Springer, 2003.

\bibitem{Boas54}
R.~P.~Jr. Boas,
\newblock {\em Entire functions},
\newblock Acacemic Press, 1954.

\bibitem{Brillinger01}
David~R. Brillinger,
\newblock {\em Time series: data analysis and theory},
\newblock Classics in Applied Mathematics. SIAM, 2001.

\bibitem{Kaveh86}
Mostafa Kaveh and Arthur~J. Barabell,
\newblock ``The statistical performance of the {M}{U}{S}{I}{C} and the
  {M}inimum-{N}orm algorithms in resolving plane waves in noise,''
\newblock {\em IEEE Transactions of Acoustics, Speech, and Signal Processing},
  vol. ASSP-34, no. 2, pp. 331--341, Apr 1986.

\bibitem{Tuncer09}
Engin Tuncer and Benjamin Friedlander, Eds.,
\newblock {\em Classical and modern direction-of-arrival estimation},
\newblock Elsevier, 2009.

\bibitem{Shaw16}
A.~K. Shaw,
\newblock ``Improved wideband {D}{O}{A} estimation using modified {T}{O}{P}{S}
  (m{T}{O}{P}{S}) algorithm,''
\newblock {\em IEEE Signal Processing Letters}, vol. 23, no. 12, pp.
  1697--1701, Dec 2016.

\bibitem{Mason02}
John~C. Mason and David~C. Handscomb,
\newblock {\em Chebyshev polynomials},
\newblock CRC Press, 2002.

\bibitem{Stewart90}
G.~W. Stewart and Ji~guang Sun,
\newblock {\em Matrix Perturbation Theory},
\newblock Academic Press, Inc., 1990.

\bibitem{Janssen88}
P.H.M. Janssen and P.~Stoica,
\newblock ``On the expectation of the product of four matrix-valued {G}aussian
  random variables,''
\newblock {\em IEEE Transactions on Automatic Control}, vol. 33, no. 9, pp.
  867--870, 1988.

\end{thebibliography}

\end{document}